%% file: BP_gains.tex
\documentclass[twocolumn]{aa}

\usepackage{graphicx}
\usepackage{amsmath,amsfonts,amssymb}
\usepackage{txfonts}
\usepackage{color}
\usepackage{natbib}
\usepackage{caption,subcaption}
\usepackage{float}
\usepackage{dblfloatfix}
\usepackage{afterpage}
\usepackage{ifthen}
\usepackage[morefloats=12]{morefloats}
\usepackage{placeins}
\usepackage{multicol}
\bibpunct{(}{)}{;}{a}{}{,}
\usepackage[switch]{lineno}
\definecolor{linkcolor}{rgb}{0.6,0,0}
\definecolor{citecolor}{rgb}{0,0,0.75}
\definecolor{urlcolor}{rgb}{0.12,0.46,0.7}
\usepackage[breaklinks, colorlinks, urlcolor=urlcolor,
    linkcolor=linkcolor,citecolor=citecolor,pdfencoding=auto]{hyperref}
\hypersetup{linktocpage}
\usepackage{bold-extra}

\input{Planck}

\def\WMAP{\textit{WMAP}}
\def\COBE{\textit{COBE}}
\def\LCDM{$\Lambda$CDM}

\def\commander{\texttt{Commander}}

\renewcommand{\d}[0]{\vec{d}}

\newcommand{\n}[0]{\vec{n}}

\newcommand{\s}[0]{\vec{s}}
\renewcommand{\a}[0]{\vec{a}}

\newcommand{\B}[0]{\tens{B}}
\newcommand{\T}[0]{\tens{T}}
\newcommand{\tG}[0]{\tens{G}}

\renewcommand{\L}[0]{\tens{L}}
\newcommand{\g}[0]{\vec{g}}
\newcommand{\bgamma}[0]{\boldsymbol{\gamma}}
\newcommand{\N}[0]{\tens{N}}

\renewcommand{\S}[0]{\tens{S}}
\renewcommand{\r}[0]{\vec{r}}

\renewcommand{\v}[0]{\vec{v}}
\renewcommand{\P}[0]{\tens{P}}

\newcommand{\BP}{\textsc{BeyondPlanck}}
\newcommand{\lfi}[0]{LFI}
\newcommand{\hfi}[0]{HFI}
\newcommand{\npipe}[0]{\texttt{NPIPE}}

\newcommand{\srollTwo}[0]{\texttt{SROLL2}}
\newcommand{\ti}[0]{_{t, i}}
\newcommand{\qi}[0]{_{q, i}}

\newcommand{\x}[0]{\vec{x}}
\newcommand{\tot}[0]{^{\mathrm{tot}}}
\newcommand{\corr}[0]{^{\mathrm{corr}}}
\newcommand{\wn}[0]{^{\mathrm{wn}}}
\newcommand{\sky}[0]{^{\mathrm{sky}}}
\newcommand{\orb}[0]{^{\mathrm{orb}}}

\def\inv{^{-1}}

\begin{document}

\title{\bfseries{\scshape{BeyondPlanck}} VII. Bayesian estimation of gain and absolute calibration for CMB experiments}
\input{authors_07.tex}
\authorrunning{BeyondPlanck Collaboration}
\titlerunning{Calibration}

\abstract{We present a Bayesian calibration algorithm for CMB observations as implemented within the global end-to-end \BP\ framework, and apply this to the \Planck\ Low Frequency Instrument (\lfi) data. Following the most recent \Planck\ analysis, we decompose the full time-dependent gain into a sum of three nearly orthogonal components: One absolute calibration term, common to all detectors; one time-independent term that can vary between detectors; and one time-dependent component that is allowed to vary between one-hour pointing periods.  Each term is then sampled conditionally on all other parameters in the global signal model through Gibbs sampling. The absolute calibration is sampled using only the orbital dipole as a reference source, while the two relative gain components are sampled using the full sky signal, including the orbital and Solar CMB dipoles, CMB fluctuations, and foreground contributions. We discuss various aspects of the data that influence gain estimation, including the dipole/polarization quadrupole degeneracy and processing masks. Comparing our solution to previous pipelines, we find good agreement in general, with relative deviations of $-0.67\,\%$ ($-0.84\,\%$) for 30\,GHz, $0.12\,\%$ ($-0.04\,\%$) for 44\,GHz and $-0.03\,\%$ ($-0.64\,\%$) for 70\,GHz, compared to \Planck\ DR4 (Planck 2018).  We note that the \BP\ calibration is performed globally, which results in better inter-frequency consistency than previous estimates. Additionally, \WMAP\ observations are used actively in the \BP\ analysis, and this both breaks internal degeneracies in the \Planck\ data set and results in better overall agreement with \WMAP. 
Finally, we use a Wiener filtering approach to smoothing the gain estimates. We show that this method avoids artifacts in the correlated noise maps due to over-smoothing the gain solution, which is difficult to avoid with methods like boxcar smoothing, as Wiener filtering by construction maintains a balance between data fidelity and prior knowledge. Although our presentation and algorithm are currently oriented toward \lfi\ processing, the general procedure is fully generalizable to other experiments, as long as the Solar dipole signal is available to be used for calibration.}


\maketitle

\section{Introduction}
\label{sec:introduction}
The cosmic microwave background (CMB) anisotropies are among the most important observables available to cosmologists, and accurate determination of their statistical properties has been a main goal for a multitude of collaborations and experiments during the last three decades \citep[e.g.,][and references therein]{smoot:1992,debernardis:2000,kovac:2002,bennett2012,planck2016-l01}. The \BP\ project \citep{bp01} is an initiative aiming to establish a common multi-experiment analysis platform that supports global end-to-end Bayesian analysis of raw time-ordered data (TOD) produced by such experiments, as well as seamless propagation of low-level uncertainties into all high-level products, including frequency and component maps, the CMB angular power spectra, and cosmological parameters. As a first demonstration, we apply this framework to the \Planck\ LFI data, as presented in this and a suite of companion papers.

A fundamentally important step in any CMB analysis pipeline is photometric calibration---the process of mapping the instrument readout to the incoming physical signal. In general, this procedure involves comparing some specific feature in the measured data with a known calibration model, for instance the CMB dipole or astrophysical foreground signal (or both), or by comparing the total measured power with a reference load with a known physical temperature \citep[e.g.,][]{planck2014-a06}. 

The multiplicative factor that converts between sky signal and detector readout is called the gain. This factor typically depends on the local environment of the detectors, such as the ambient temperature, and is therefore  in principle different for each sample. However, as long as the detectors are thermally stable on reasonably long time scales, it is usually a good approximation to assume that the gain is constant over some short period of time, or at least that it is smoothly varying in time. For instance, the \WMAP\ team adopted and fitted a six-parameter model for the gain, using housekeeping data such as focal plane temperature measurements to interpolate in time \citep{jarosik2007, Hinshaw_2009, wmapexsupp}. For \lfi, we will assume that the gain factor is constant throughout each \Planck\ pointing period (PID) -- the timescale during which the satellite scans a given ``circle'' on the sky; these last roughly an hour each. We will also assume that the gain is varying smoothly between neighboring PIDs, except during a small set of events during which the instrument was actively modified by the mission control center, for instance during cooler maintenance. 

The \Planck\ \lfi\ Data Processing Centre (DPC) \citep{planck2013-p02,planck2014-a06,planck2016-l02} used an onboard 4\,K reference load to support the 30\,GHz calibration for the early results, while for the other channels, and for all channels in later releases, they relied primarily on the CMB dipole signal. Gain fluctuations and correlations were modelled and suppressed by boxcar averaging over a signal-to-noise dependent window size. The \Planck\ \hfi\ DPC \citep{planck2013-p03f,planck2014-a09,planck2016-l03} also used the CMB dipole signal for gain estimation, but in this case they assumed a constant gain factor throughout the whole mission, relying on the excellent thermal stability of the \Planck\ instrument. Apparent gain variations were instead assumed to arise from non-linearities in the analog-to-digital conversion module, which then allowed for a deterministic correction. A similar approach has also been adopted by the recent \srollTwo\ re-analysis initiative \citep{delouis:2019}. We refer to the latest DPC maps as \Planck\ 2018.

In the most recent official analysis (\Planck\ DR4\footnote{Sometimes referred to as \npipe.}; \citealp{planck2020-LVII}), the \Planck\ team adopted the LFI gain model for all channels up to and including the 857\,GHz channel. A novelty introduced in that analysis, however, was a decomposition of the gain factor into two nearly orthogonal components: an absolute (or baseline) gain factor, which was assumed to be constant for the entire mission, and a detector-specific gain mismatch factor that could vary both in time and between detectors. This approach allowed estimation of each component separately, using calibrators that are better suited to each component. For example, the low signal-to-noise (but well-understood) orbital dipole was used to calibrate the absolute gain factor due to the long integration time involved in estimating this particular component. Solving for all relevant factors was then performed jointly with other relevant quantities.

In this paper, we adopt the \Planck\ DR4 approach, and decompose the full gain into the above-mentioned components, and we estimate these jointly with all other parameters in the full model. Thus, the main novel feature presented in this paper is the integration of the gain estimation procedure within a larger Gibbs framework, as summarized by \citet{bp01}, which performs joint estimation of all relevant parameters in a statistically consistent manner, including the CMB and astrophysical foreground sky signal.

The rest of the paper is structured as follows: In Sect.~\ref{sec:intuition}, we aim to build intuition regarding gain estimation, presenting the general data model that we use and highlighting various important features of this model, as applied to real-world \lfi\ observations. Next, in Sect.~\ref{sec:methodology} we describe our main gain estimation procedure, before showing results in Sect.~\ref{sec:results}, and comparing these with those derived by other pipelines. Finally, we summarize in Sect.~\ref{sec:conclusions}, with an eye toward future experiments and applications.

\section{Data and modelling considerations}
\label{sec:intuition}

We start our presentation with a general discussion of the
gain-related data model, and how to account for various complications
that arise when fitting this to real-world data.

\subsection{Data model}
\label{sec:gain_modelling}
As described by \citet{bp01}, the main goal of the \BP\ analysis framework is to develop an end-to-end Bayesian analysis platform for CMB data, starting from raw time-ordered data. As for most Bayesian problems, the key step in our approach is therefore to write down an explicit parametric model for the observed data from a given detector, $d\ti$, where $t$ is the index denoting the sample,\footnote{Here, a ``sample'' means the detector readout at every $1/f_\mathrm{samp}$ seconds, where $f_\mathrm{samp}$ is the sampling frequency of the instrument. The whole set of these samples is called the time-ordered data (TOD). The sampling frequency for the three LFI instruments are 32.5\,Hz, 46.5\,Hz, and 78.8\,Hz for the 30\,GHz, 44\,GHz, and 70\,GHz instrument, respectively.} and $i$ is the index denoting the detector in question. In the current analysis, we adopt the following high-level model,
\begin{equation}
    d\ti = g\ti s\ti\tot + n\ti\corr + n\ti\wn,
    \label{eq:gen_data_model}
\end{equation}
where $n\ti\corr$ and $n\ti\wn$ are correlated and white noise, respectively, $g\ti$ is the gain factor, and $s\ti\tot$ denotes the total signal. Here, $s\ti\tot$ is given in kelvin, while $d\ti$ is the instrument readout, which is measured in volts, meaning that the unit for $g\ti$ becomes $[\mathrm{V/K}]$. The total signal can further be decomposed into
\begin{align}
    s\ti\tot &= s\ti\sky + s\ti\orb + s\ti^{\mathrm{fsl}} \nonumber \\ 
    &= P_{tp, i}\B^{\mathrm{symm}}_{pp', i} s_{p'}\sky + P_{tp, i}\B^{\mathrm{asymm}}_{pp', i}s_{p'}\orb + P_{tp, i}\B^{\mathrm{asymm}}_{pp', i}s_{p'}^{\mathrm{fsl}} \nonumber \\
    & = P_{tp, i}\left[\B^{\mathrm{symm}}_{pp', i} s_{p'}^{\mathrm{sky}} + \B_{pp', i}^{\mathrm{asymm}}\left(s_{p'}\orb + s_{p'}^{\mathrm{fsl}}\right)\right].
\end{align}
In this expression, \P\ is the pointing matrix, which contains the mapping between the pointing direction the instrument, $p$ and the sample index $t$; $\B^{\mathrm{symm}}$ and $\B^{\mathrm{asymm}}$ denote convolution with the symmetric and asymmetric beams, which quantify the fraction of the total signal coming from direction $p'$ when the instrument is pointing towards $p$; $s^{\mathrm{sky}}$ is the sky signal (including the Solar dipole); $s^{\mathrm{orb}}$ is the orbital dipole (to be discussed below); and $s^{\mathrm{fsl}}$ represents signal leakage through the far-sidelobes. For further details regarding any of these objects, we refer the interested reader to \citet{bp01} and references therein.

The main topic of the current paper is estimating $g\ti$. In this respect, it is important to note that all other free parameters in the data model, including $s\ti\sky$ and $n\ti\corr$, are also unknown, and must be estimated jointly with $g\ti$. Casting this statement into Bayesian terms, we wish to sample from the \emph{posterior distribution},\footnote{Here, and elsewhere, boldface quantities generally mean vectors. Which vector space they belong to will to a large degree be evident from the subscripts -- in this case, there are no subscripts, meaning that the vectors contain all samples from all detectors.}
\begin{equation}
    P(\g, \s\tot, \s\orb, \n\corr, \ldots \mid \d).
\end{equation}
That is, we aim to model the \emph{global} state of the instrument and data, and map out the probability of various points in parameter space by sampling from this distribution. This may at first glance seem like a intractable problem. However, a central component of the \BP\ framework is parameter estimation through \emph{Gibbs sampling}. According to the theory of Gibbs sampling, samples from a joint posterior distribution may be drawn by iteratively sampling from all \emph{conditional} distributions. In other words, when sampling the gain, we may assume that the sky signal and correlated noise parameters are perfectly known. And likewise, when sampling the sky signal or correlated noise parameters, we may assume that the gain is perfectly known. The correlations between these various parameters are then probed by performing hundreds or thousands of iterations of this type. 

Thus, for the purposes of calibration alone, we do not need to be
concerned with many aspects that indirectly affect the gain, such as
CMB dipole or correlated noise estimation
\citep{planck2013-p02,planck2014-a06,planck2016-l02}. Instead, we are
here concerned only with defining an adequate model for $\g$, and
expressing this in a way that minimizes degeneracies with parameters
in the Gibbs chain. 

As discussed above, the gain is generally not constant in time. A very conservative (and somewhat na\"ive) model would therefore be to assume that the gain is in fact different for every sample $t$. However, this model clearly does not take into account our full knowledge about the instrument \citep{planck2013-p05}. In particular, we do know that the gain is expected to correlate with the detector temperature, and this temperature does not change significantly on timescales of just one sample. Rather, based on available housekeeping data, a good assumption is that the gain is constant within a given pointing period (PID, or \emph{scan}) -- which is defined as a collection of samples measured over a period of about an hour, during which the instrument spins about its axis once per minute while keeping the spin axis vector stationary. Between each scan, the instrument performs a slight adjustment of the satellite spin axis, ensuring that new sky areas are covered in consecutive pointing periods. 

To reflect the assumption of constant gain within each scan, we rewrite our data model as follows,
\begin{equation}
    d\ti = g\qi s\ti\tot + n\ti\corr + n\ti\wn,
    \label{eq:gen_data_model_scan}
\end{equation}
where $q$ now denotes PID. Thus, $t$ is used to indicate a specific sample, while $q$ represents a collection of samples.

From Eq.~\eqref{eq:gen_data_model_scan}, we immediately note the presence of two important degeneracies, involving the sky signal and noise, respectively. If we attempt to fit for $g$, $s\tot$, and $n\corr$ simultaneously, without knowing anything about any of them, we see that a given solution, say, $\{g_0, s_0\tot, n_0\corr\}$, will result in an identical goodness-of-fit as another solution $\{g_1, s_1\tot, n_1\corr\}$, as long as either
\begin{equation}
  g_1 = g_0\frac{s_0\tot}{s_1\tot},
\end{equation}
or
\begin{equation}
  n_1\corr = n_0\corr + g_0s_0\tot - g_1 s_1\tot.
\end{equation}
In other words, the gain is multiplicatively degenerate with the
signal, and additively degenerate with the correlated noise. Such
degeneracies are mainly a computational problem, since with two degenerate parameters in a Gibbs chain, exploring the resulting distributions takes a much larger number of samples than for uncorrelated parameters. A main topic of this
paper is how to break these degeneracies in a statistically
self-consistent and computationally efficient manner.

\subsection{Absolute versus relative gain calibration}
\label{sec:absrelgain}
So far, we have been talking about the calibration of a given detector in isolation, which relates to what we call \emph{absolute} calibration. Absolute calibration refers to correctly determining the ``true'' value of the gain, and is important for accurately determining the emitted intensity of astrophysical components, such as the CMB.

Another closely related concept is \emph{relative} calibration,\footnote{Note that our definition differs slightly from the \Planck\ 2018 definition of relative calibration. In their nomenclature, relative calibration refers to temporal fluctuations of the gain around the mean within a given detector.} which quantifies calibration differences between radiometers. Because of \Planck 's scanning strategy, which only provides weak cross-linking\footnote{The scanning strategy adopted by \Planck\ means that the time interval between successive measurements of the same point on the sky with the same detector from a different angle can be as much as six months (if the point lies along the ecliptic). During this time, several environmental parameters of the detector may have changed.} \citep{planck2011-1.1}, it is impossible to estimate the three relevant Stokes parameters (the intensity, $I$, and two linear polarization parameters, $Q$ and $U$) independently for each detector. Rather, the polarization signal is effectively determined by considering pairwise differences between detector measurements, while properly accounting for their relative polarization angle differences at any given time. Therefore, any instrument characterization error that induces spurious detector differences will be partially interpreted by the analysis pipeline as a polarization signal. If our relative gain calibration is wrong, such differences will be introduced.

Given the high sensitivity of current and future CMB experiments, the gain must be estimated to a fractional precision better than $\mathcal{O}(10^{-3})$ for robust CMB temperature analysis, and better than $\mathcal{O}(10^{-4})$ for robust polarization analysis. Accurate relative calibration is thus even more important than accurate absolute calibration, and this will, as discussed in the next section, inform the choices we make on how to estimate these two types of calibration.

\subsection{The Solar and orbital CMB dipoles}
\label{sec:dipoles}
One of the most powerful ways to break the signal/gain degeneracy mentioned in the previous section is to observe a source of known brightness. If that source happens to be significantly stronger than other sources in the same area of the sky, we could fix $s\ti\tot$ in Eq.~\eqref{eq:gen_data_model_scan} and the gain would essentially be determined by the ratio of the data to the known source brightness. 

\begin{figure}[t]
  \center
  \includegraphics[width=\linewidth]{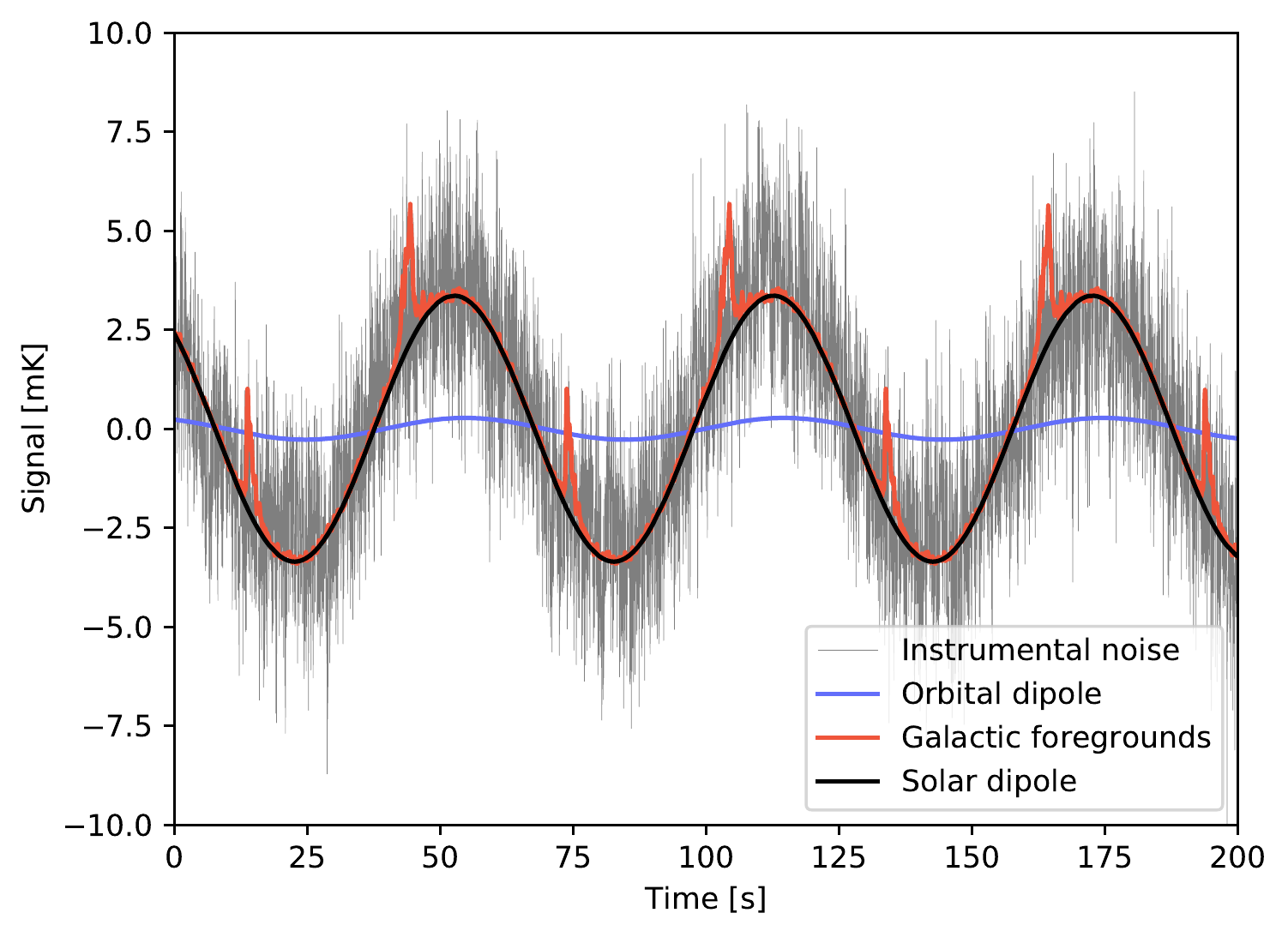}
    \caption{Comparison of different contributions to the time-ordered
      data seen by \Planck\ at 30\,GHz, for a PID whose orientation is close to perpendicular to the dipole axis. The blue and black curves
      show the orbital and Solar CMB
      dipoles, respectively, while the red line shows contributions
      from small-scale CMB fluctuations and Galactic foregrounds. The
      gray line shows instrumental noise. }
  \label{fig:dipole_strengths}
\end{figure}

Unfortunately, the number of available astrophysical calibration
sources that may be useful for CMB calibration purposes is very
limited, given the stringent requirements discussed in
Sect.~\ref{sec:absrelgain}. For instance, the brightness temperature
of individual planets within the Solar system is only known to about
5\,\% \citep{planck2014-a09}, while few other local sources are known
with a precision better than 1\,\%.

The key exception is the CMB dipole. The peculiar velocity of the
\Planck\ satellite relative to the CMB rest frame induces a strong
apparent dipole on the sky due to the Doppler effect. Specifically,
photons having an anti-parallel velocity relative to the satellite
motion are effectively blue-shifted, while photons with a parallel
velocity are redshifted.

It is useful to decompose the peculiar spacecraft velocity into two
components; the motion of the Solar system relative to the CMB
rest frame, $\vec{v}_{\mathrm{solar}}$, and the orbital motion of the
\Planck\ satellite relative to the Solar system barycenter,
$\vec{v}_{\mathrm{orbital}}$. Thus, the total velocity of the
satellite relative to the CMB rest frame is $\vec{v}_{\mathrm{tot}} =
\vec{v}_{\mathrm{solar}} + \vec{v}_{\mathrm{orbital}}$. Taking into
account the full expression for the relativistic Doppler effect, the
induced dipole reads
\begin{equation}
    s^{\mathrm{dip}}(\vec{x}, t) = T_{\mathrm{CMB}}\left(
    \frac{1}{\gamma(t)(1 - \vec{\beta}(t) \cdot \vec{x})} - 1 \right),
    \label{eq:dipole}
\end{equation}
where $\vec{\beta} = \v_{\mathrm{tot}}/c$, and $\gamma = (1-|\vec\beta|^2)^{-1/2}$. The total
dipole is time dependent because of the motion of the satellite over
the course of the mission. We can similarly define a Solar dipole,
$s^{\mathrm{solar}}(\vec{x})$ and an orbital dipole,
$s^{\mathrm{orb}}(\vec{x}, t)$, which are induced by only the Solar and
orbital velocities alone, respectively.
\begin{figure}[t]
  \center
  \includegraphics[width=\linewidth]{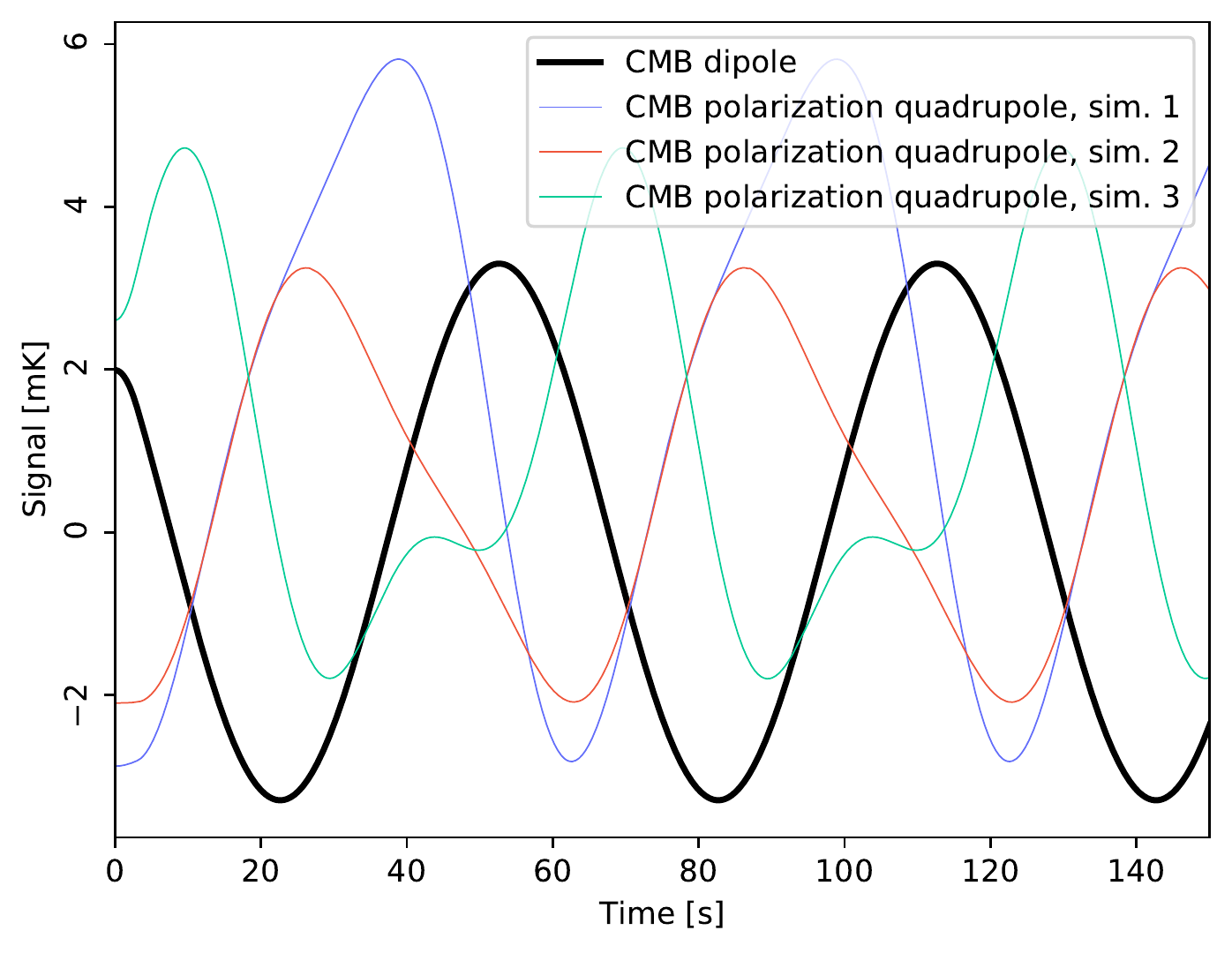}
    \caption{Comparison of the CMB temperature dipole (thick black
      line) observed through the \Planck\ scanning strategy with three
      random polarization quadrupole simulations (thin colored lines);
      the latter have been scaled by a factor of $10^{4}$ for
      visualization purposes. 
    }
  \label{fig:dip_quadr_degeneracy}
\end{figure}
Both dipoles play crucial roles in CMB calibration; the orbital dipole
for absolute calibration and the Solar dipole for relative
calibration.

Starting with the orbital dipole, we note that this
depends only on the satellite's velocity with respect to the Sun. This
is exceedingly well measured through radar observations, and known
with a precision better than $1\,\textrm{cm}\,\textrm{s}^{-1}$
\citep{godard2009}. For an orbital speed of
$30\,\mathrm{km}\,\mathrm{s}^{-1}$, this results in an overall
relative precision better than $\mathcal{O}(10^{-6})$. However,
Eq.~\eqref{eq:dipole} also depends on the CMB monopole, which is
measured by \COBE-FIRAS to $2.72548\,\textrm{K} \pm 0.57\,\textrm{mK}$
\citep{fixsen2009}, corresponding to a relative uncertainty of
0.02\,\% or $\mathcal{O}(10^{-4})$. Thus, the absolute calibration of
any current and future CMB experiment cannot be determined with a
higher absolute precision than $\mathcal{O}(10^{-4})$ until a
next-generation CMB spectral distortion experiment, for instance PIXIE
\citep{kogut:2011}, is launched. Still, this precision is more than
sufficient for \Planck\ calibration purposes.

The second CMB dipole component corresponds to the Sun's motion with
respect to the CMB rest frame. While this velocity is intrinsically
unknown, one may estimate this from the relative amplitude of the
Solar and orbital dipoles. This is illustrated in
Fig.~\ref{fig:dipole_strengths}, which compares the orbital and Solar
dipole signals (blue and black curves) with contributions from
Galactic foreground emission and instrumental noise at 30\,GHz for
about three minutes of time-ordered observations. The Solar dipole is
effectively determined by the relative amplitude ratio between the
black and blue curves in this figure.

\begin{figure*}[t]
    \centering
    \begin{subfigure}{0.45\textwidth}
       \includegraphics[width=0.95\linewidth]{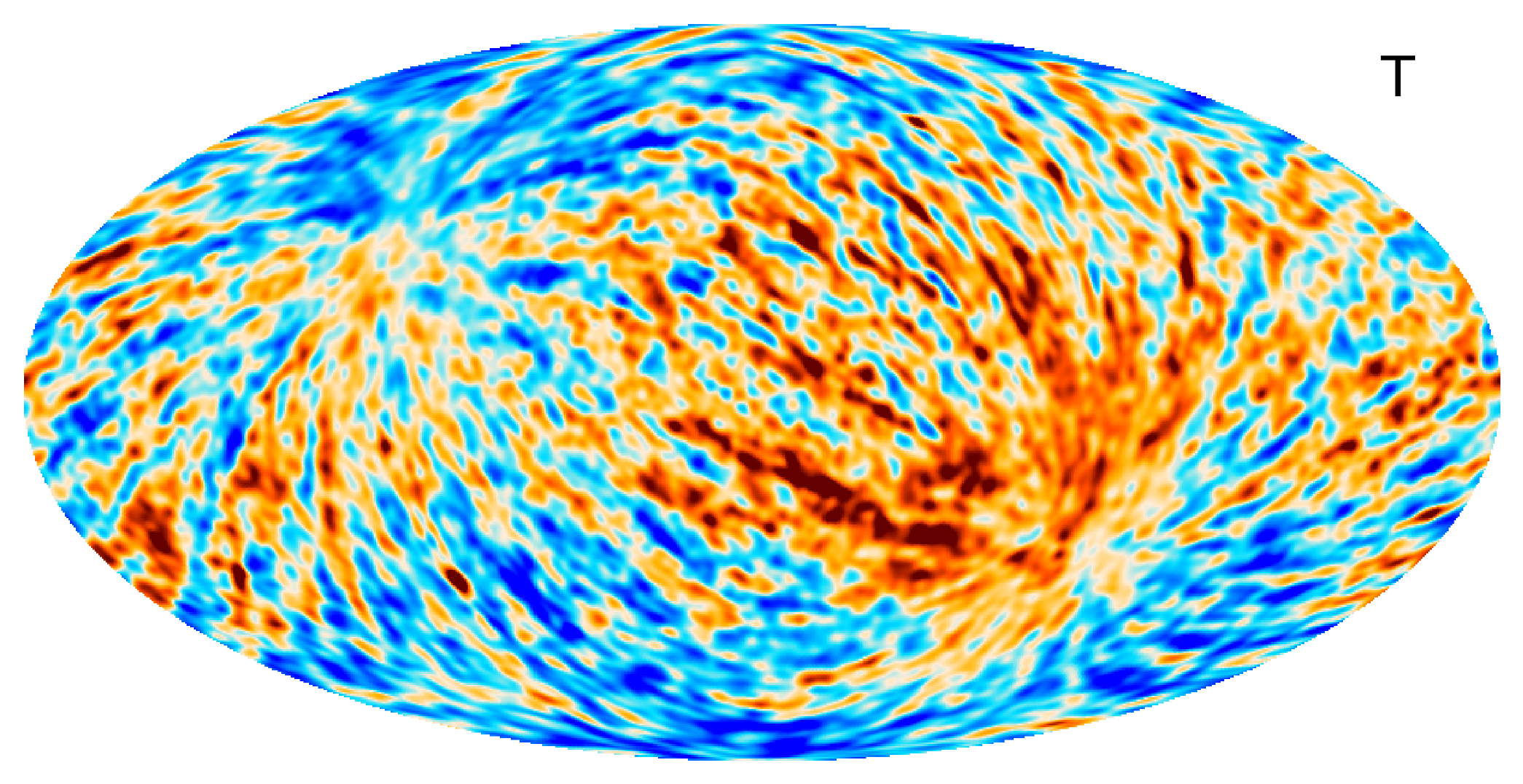}
    \end{subfigure}
    \begin{subfigure}{0.45\textwidth}
       \includegraphics[width=0.95\linewidth]{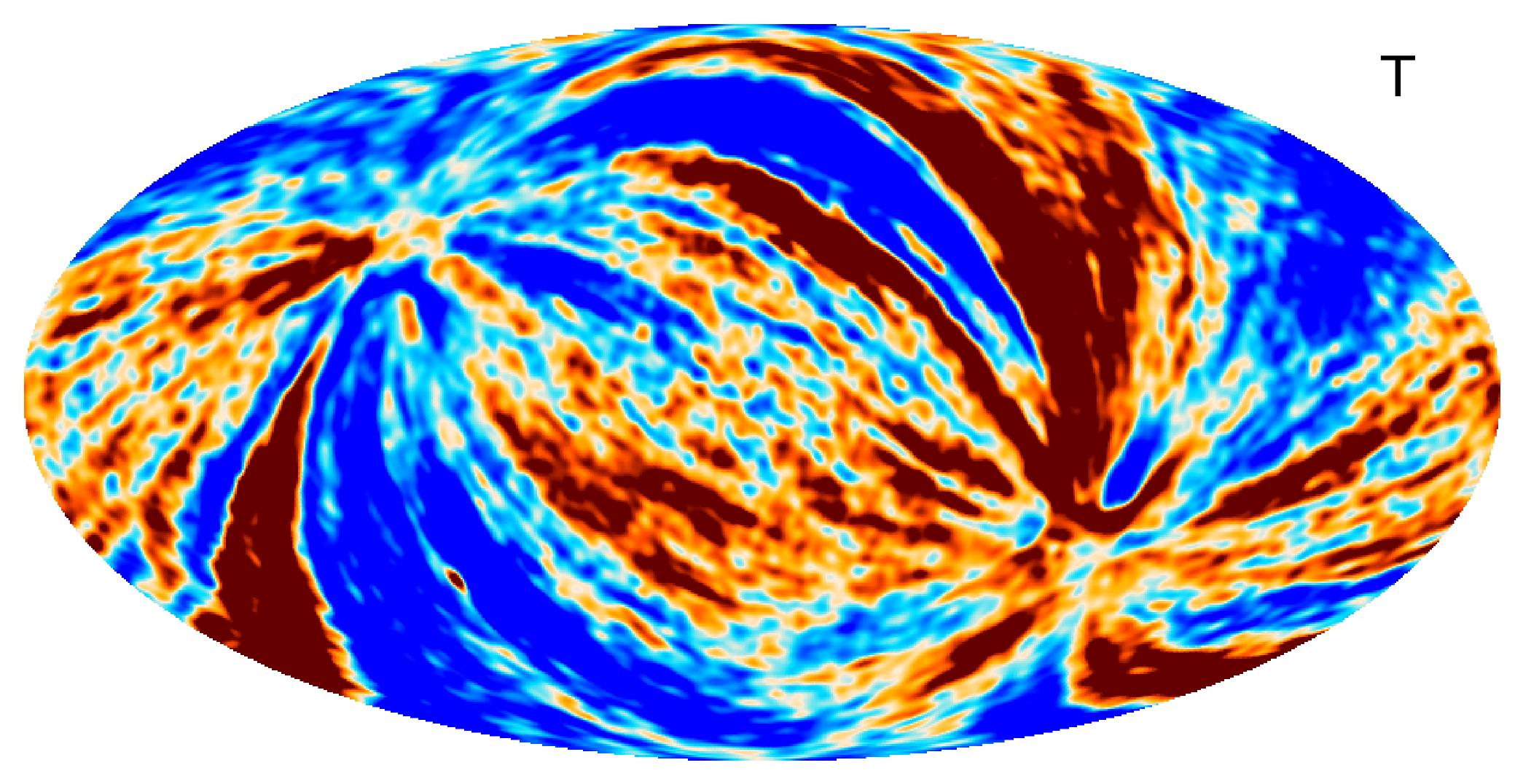}
    \end{subfigure}
    \begin{subfigure}{0.45\textwidth}
       \includegraphics[width=0.95\linewidth]{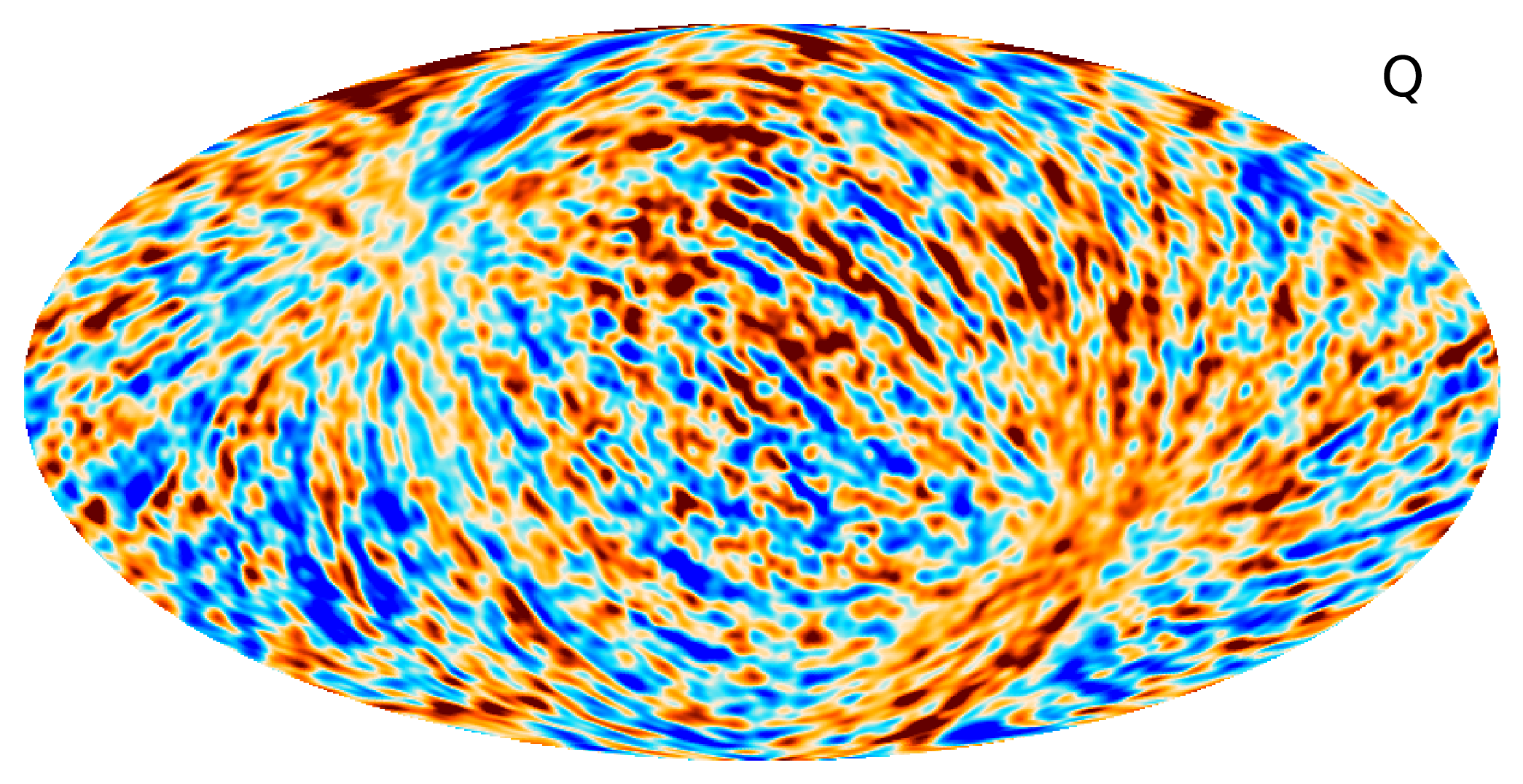}
    \end{subfigure}
    \begin{subfigure}{0.45\textwidth}
       \includegraphics[width=0.95\linewidth]{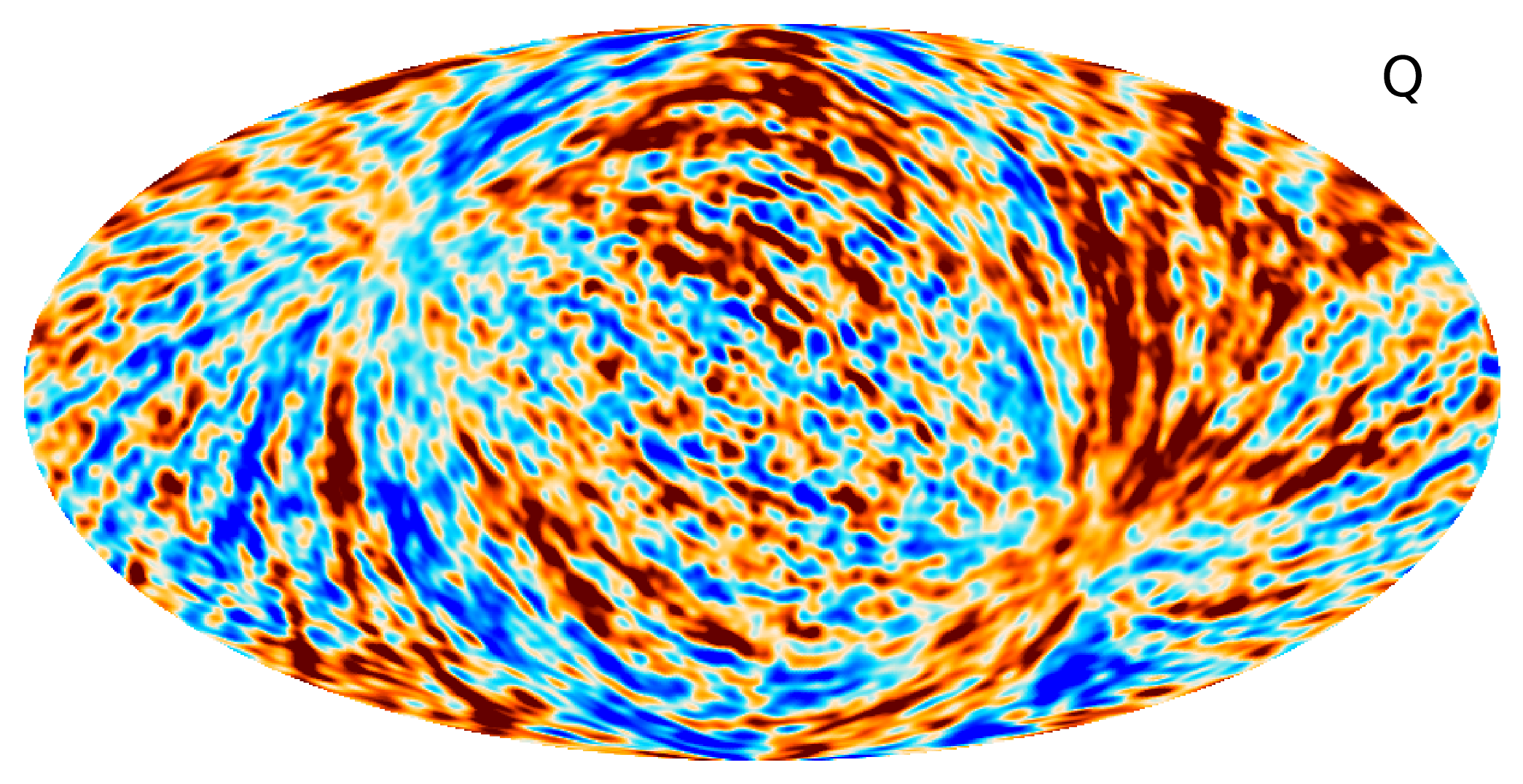}
    \end{subfigure}
    \begin{subfigure}{0.45\textwidth}
       \includegraphics[width=0.95\linewidth]{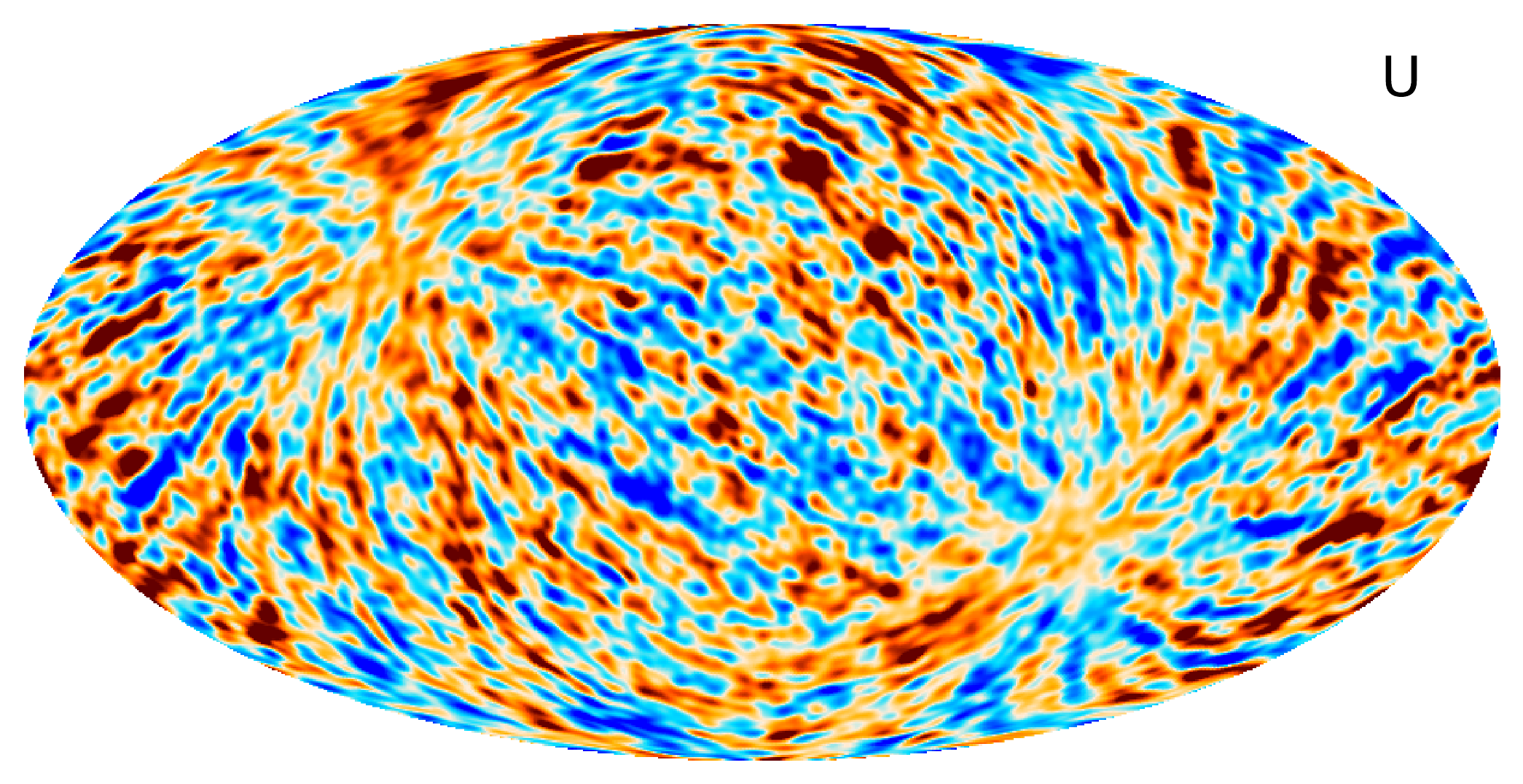}
    \end{subfigure}
    \begin{subfigure}{0.45\textwidth}
       \includegraphics[width=0.95\linewidth]{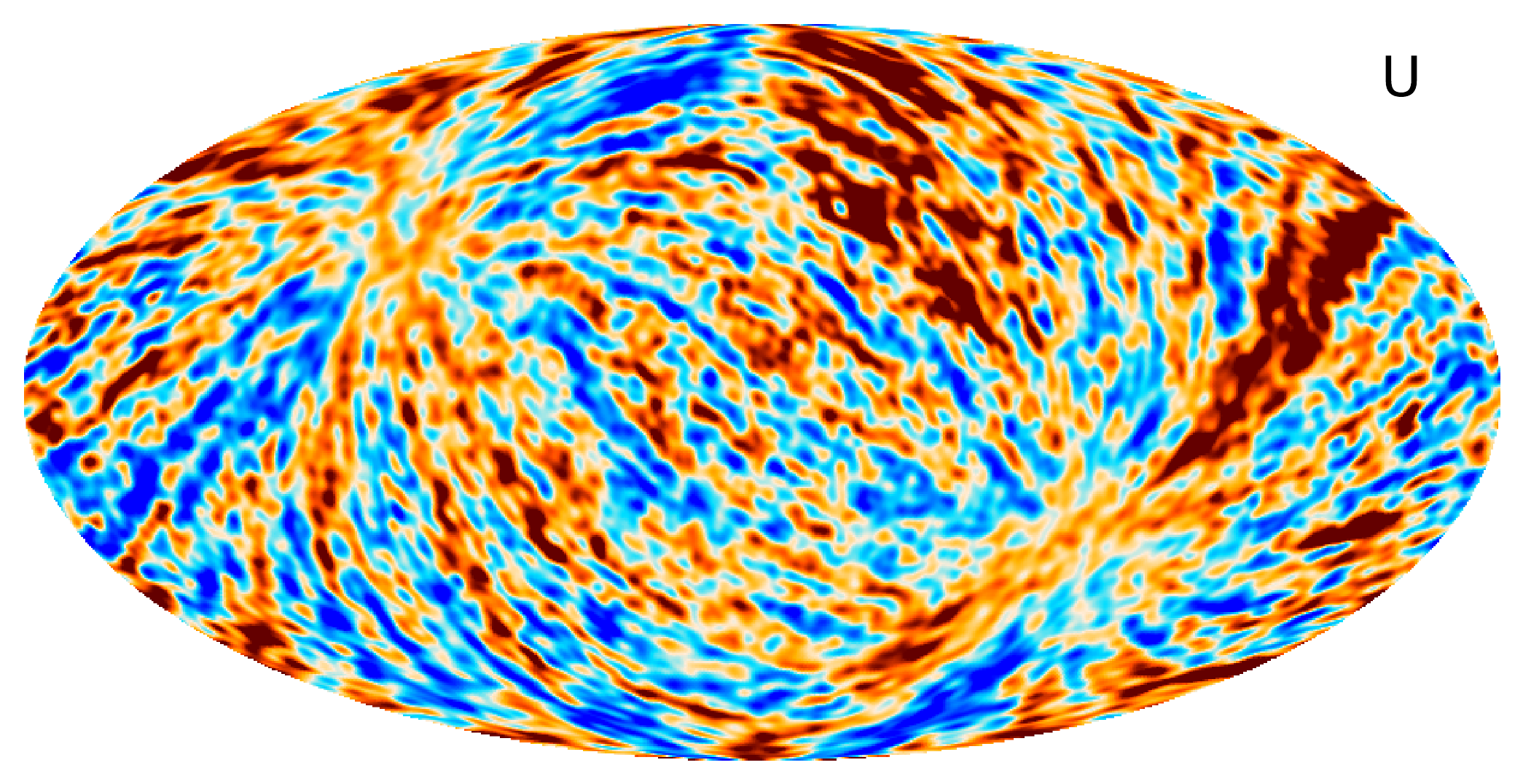}
    \end{subfigure}
    \begin{subfigure}{0.45\textwidth}
       \includegraphics[width=0.95\linewidth]{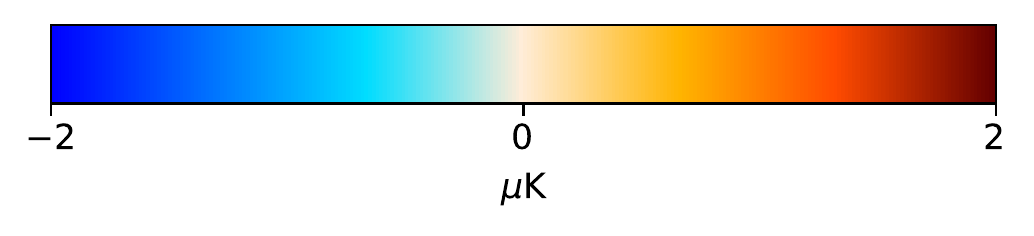}
    \end{subfigure}
    \caption{Correlated noise maps for the 30\,GHz channel in a Gibbs chain that includes (\emph{left panel}) or neglects (\emph{right panel}) gain time-dependencies. All maps are smoothed to a common resolution of 2.5$^\circ$ FWHM.}
    \label{fig:ncorr_timedep}
\end{figure*}

Based on this approach, the most recent \Planck\ analyses have
determined that the Solar CMB dipole amplitude is about 3.36\,mK,
corresponding to Solar velocity of about
370\,$\textrm{km}\,\textrm{s}^{-1}$ \citep{planck2016-l01,planck2020-LVII}. For
comparison, large-scale CMB polarization fluctuations typically
exhibit variations smaller than $\mathcal{O}(1\muK)$
\citep{planck2016-l04}, and consequently the relative calibration of
different detectors must be better than $\mathcal{O}(10^{-4})$ to
avoid significant contamination of the polarization signal by the Solar CMB
dipole. Achieving this level of precision in the presence of correlated noise,
Galactic foregrounds, far sidelobe contamination and other sources of
systematic uncertainties is the single most difficult challenge associated with
large-scale CMB polarization science.

\subsection{The degeneracy between the CMB temperature dipole and
  polarization quadrupole}
\label{sec:quadrupole}

In Sect.~\ref{sec:gain_modelling} we noted that the gain is multiplicatively
degenerate with the signal, and additively degenerate with the
noise. Within this broad categorization, there are also some
particularly important gain/signal degeneracies that are worth highlighting, and
perhaps the most prominent example is that with respect to the CMB
polarization quadrupole. To illustrate this, consider the case in
which two detectors report different CMB dipole amplitude signals; how could such
a difference be explained? One possible explanation is a calibration
mismatch, i.e., that the absolute calibration of one or both detectors
is mis-estimated.

Another possible explanation could be a large polarization
CMB \emph{quadrupole} signal. Due to the scanning strategy adopted by
\Planck, in which the same ring is observed repeatedly for one hour, a
polarization quadrupole can easily appear with a dipolar signature,
depending on the particular phase orientation of the mode in
question. This is illustrated in
Fig.~\ref{fig:dip_quadr_degeneracy}. The black thick line shows the
CMB temperature dipole as a function of time, while the colored thin
lines show three random polarization quadrupole simulations, all
observed with the \Planck\ scanning strategy. Out of the three random
quadrupole simulations, two have a time-domain behaviour that
very closely mimics the CMB temperature dipole, and in the presence of
noise and instrumental effects, it would be exceedingly difficult
to distinguish between the two models.

This is a perfect recipe for a degenerate system, and one
that carries the potential of contaminating any large-scale
polarization reconstruction. It is, however, important to note that
this particular degeneracy appears with a very specific morphology,
and affects only a handful of spatial polarization modes, as defined
by projecting the CMB Solar dipole onto the \Planck\ scanning
strategy. Recognizing the importance of this degeneracy, previous
\Planck\ analyses have adopted different strategies of resolving the
issue. For instance, both the \Planck\ 2018 and \Planck\ DR4 analyses
have opted to disregard the CMB polarization component completely
during the calibration phase
\citep{planck2014-a03,planck2016-l02,planck2020-LVII}. This may be at least
partially justified for LFI on theoretical grounds by noting that the
CMB polarization variance on large angular scales predicted by current
best-fit \LCDM\ models is $\lesssim0.1\muK^2$, which is comparable to,
or below, the overall noise. For the significantly more sensitive HFI
instrument, this assumption is not adequate, and the recent
\Planck\ DR4 analysis therefore explicitly estimates a transfer function to
account for this effect \citep{planck2020-LVII}.

In the following, we adopt a slightly different strategy: We include the CMB polarization component in the calibration procedure, using the current sample in the Gibbs chain. In order to break the abovementioned degeneracy, we replace the polarization quadrupole of the CMB map with a random value drawn from the best-fit \LCDM\ model, using a value of $D_2^{EE} = 0.0308827 \muK^2$. Thus, we marginalize over this component and propagate the uncertainties introduced by this naturally into the Gibbs samples. We expect the combined effect of this addition to be negligible, because of the low value of the \LCDM\ prediction, and as shown in \cite{bp04} the gain estimation is unaffected by this marginalization.

\subsection{Processing masks and PID selection}
\label{sec:procmasks}
The Gibbs sampling framework used by \BP\ requires an explicit
parametric model that describes CMB, foregrounds, and the
instrument. If this model turns out to be an insufficient
representation of the actual data, the Gibbs sampling framework will
attempt to fit eventual modelling errors using the parameters that are
at its disposal. Ideally, such unexplained contributions should end up
as an excess residual in $\r = \d - \g\s\tot - \n\tot$, but in
practice they often also contaminate the other model parameters, such
as the CMB. The correlated noise, $\n\corr$, is one such parameter
that, because of its relatively unconstrained and global structure,
ends up absorbing a wide range of modelling errors, as discussed by
\citet{bp06}. Furthermore, as already noted in
Sect.~\ref{sec:gain_modelling}, there is a tight degeneracy between
the correlated noise and the gain, and $\n\corr$ is therefore a
sensitive monitor for gain errors. Figure~\ref{fig:ncorr_timedep}
illustrates this in terms of one arbitrarily selected 30\,GHz
$\n\corr$ sample from the main \BP\ analysis \citep{bp01}. The left
column shows such a sample in the default model, in which the gain is
allowed to vary from PID to PID, while the right column shows the same
when enforcing a time-independent gain. While the maps in the left
column are visually dominated by scan-aligned random stripes, as
expected for $\n\corr$, the maps in the right column (in particular
the top row) show large excesses with a dipolar pattern along each
\Planck\ scanning ring. This is the archetypal signature of gain
modelling errors, and this clearly demonstrates the need for a
time-variable gain model. At the same time, there is also a clear
quadrupolar pattern in the default configuration, with a positive
excess along the Galactic plane and a negative excess near the
Galactic poles. This structure is visually consistent with a near
sidelobe modelling residual, in the sense that the Galactic foreground
signal is slightly over-smoothed compared to the prediction of the
nominal beam model, and the resulting residual is picked up by the
correlated noise component. This is not surprising, considering that
about 1\,\% of the full LFI 30\,GHz beam solid angle is unaccounted
for in the GRASP beam model \citep{planck2016-l02}, and some of this
missing power may be in the near sidelobes. Fortunately, we also see
from the same plot that the impact of this effect is modest, and
accounts for only about 1\muK\ at 30\,GHz.

\begin{figure}[t]
  \center
    \includegraphics[width=\linewidth]{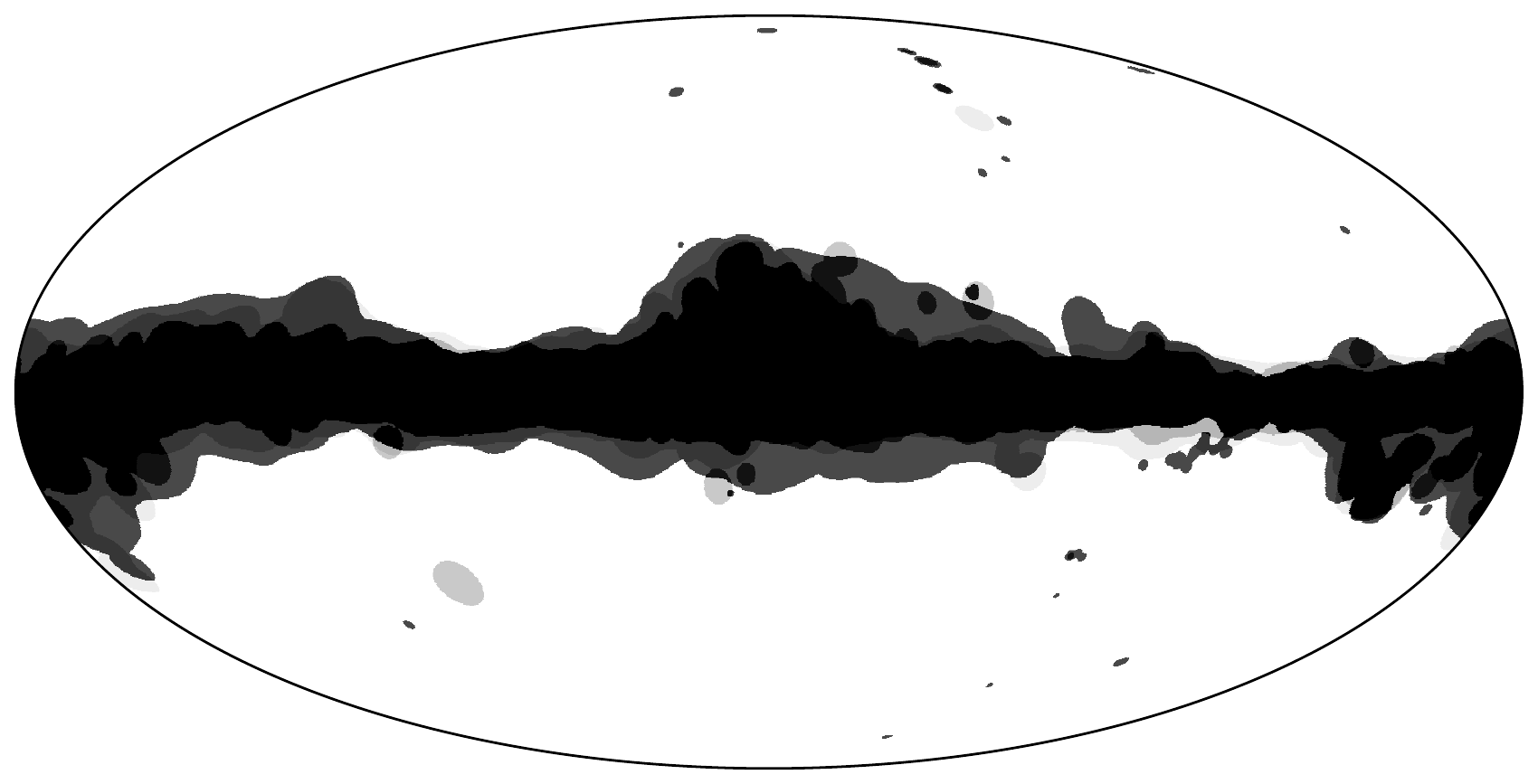}
    \caption{Processing masks used in gain and correlated noise estimation. Different shades of grey indicate different frequency masks. The allowed 30\,GHz sky fraction (light) is 0.73, the 44\,GHz sky fraction (intermediate) is 0.81, and the 70\,GHz sky fraction (dark) is 0.77. Reproduced from \citet{bp06}.}
  \label{fig:proc_mask}
\end{figure}

More generally, because we have an incomplete understanding of both
the instrument and the microwave sky, modelling errors will at some
level always be a concern when estimating both gain and correlated
noise. Furthermore, these modelling errors will typically be stronger
near the Galactic plane or bright compact sources, where foreground
uncertainties are large. For this reason, it is customary to apply a
processing mask while estimating these quantities, omitting the parts
of the sky that are least understood from the analysis. In \BP, we
define processing masks based on data-minus-signal residual maps for
each frequency \citep{bp06}, which are shown in
Fig.~\ref{fig:proc_mask}.

In addition, as discussed by \citet{bp10}, we also exclude a number of
PIDs from the main analysis, for similar reasons as for applying
processing masks. Most of these PIDs, however, do not correspond to
particularly problematic areas of the sky, but rather to unmodelled
instrumental changes or systematic errors, such as cooler maintenance or major
satellite maneuvers.  Excluded PIDs will show up as gaps in all PID plots in
this paper.

\subsection{Breaking degeneracies through multi-experiment analysis}
As described in \citet{bp01}, \BP\ includes as part of its data
selection several external data sets that are necessary to break
fundamental degeneracies within the model. One particularly important
example in this respect is the inclusion of low-resolution
\WMAP\ polarization data. In the same way that the \WMAP\ experiment
was unable to measure a few specific polarization modes on the sky due
to peculiarities in its scanning strategy \citep{jarosik2007},
\Planck\ is also unable constrain some modes as defined by its
scanning strategy \citep{planck2016-l02}. However, because the
\WMAP\ and \Planck\ scanning strategies are intrinsically very
different, their degenerate modes are not the same, and, therefore all
sky modes may be measured to high precision when analyzing both data
sets jointly.

This will be explicitly demonstrated in Sect.~\ref{sec:external},
where we compare the \BP\ sky maps to those derived individually from
each experiment. The morphology of these frequency difference maps
correspond very closely to the correction templates produced
respectively by the \WMAP\ and \Planck\ teams
\citep{jarosik2007,planck2016-l02}, and \BP\ is statistically
consistent with both data sets. Agreement is a direct and natural
consequence of performing a joint fit, and there is no need for
additional explicit template corrections for \BP.

At the same time, it is also important to note that only the
\Planck\ LFI data are currently modelled in terms of time-ordered
data, whereas the \WMAP\ sky maps and noise covariance matrices are
analyzed as provided by the \WMAP\ team. Therefore, if there should be
unknown systematics present in \WMAP, those errors will necessarily
also propagate into the various \BP\ products. An important
near-future goal is therefore to integrate also the
\WMAP\ time-ordered data into this framework. This work is already
on-going, as discussed by \citet{bp17}, but a full \WMAP\ TOD-based
analysis lies beyond the scope of the current work.

\section{Methodology}
\label{sec:methodology}
As discussed in Sect.~\ref{sec:gain_modelling}, our main goal in this
paper is to draw samples from $P(\g\mid\s\tot, \n\corr, \d, \ldots)$, the
conditional distribution of $\g$ given all other parameters. In this
section, we describe each of the various steps involved in this
process.

\subsection{Correlated noise degeneracies and computational speed-up}
\label{sec:corrnoise}

Before we present our main sampling algorithms for $\g$, we recall from
Sect.~\ref{sec:gain_modelling} that $\g$ is additively degenerate with
$\n\corr$. In a Gibbs sampling context, strong degeneracies lead to
very long Markov correlation lengths as the Gibbs sampler attempts to
explore the degenerate space between the two parameters. In order to
save computing power and time, it is therefore better to sample $\g$
and $\n\corr$ \emph{jointly}, such that for a given iteration of the
main Gibbs chain, we instead sample directly from $P(\g,
\n\corr\mid\s\tot, \d, \ldots)$.\footnote{Although this might seem somewhat
  counter-intuitive in the context of Gibbs sampling, joint sampling
  formally corresponds to re-parametrizing $\{\g, \n\corr\}$ into
  \emph{one} parameter in the Gibbs chain.}

A joint sample may be produced by means of univariate distributions
through the definition of a conditional distribution,
\begin{equation}
	P(x_1\mid x_2) \equiv \frac{P(x_1, x_2)}{P(x_2)} \Rightarrow P(x_1, x_2) = P(x_1\mid x_2)P(x_2).
\end{equation}
Thus, sampling from the joint distribution $P(\g, \n\corr\mid \s\tot, \d, \ldots)$ is equivalent to
first sampling $\g$ from its \emph{marginal} distribution with respect
to $\n\corr$, $P(\g\mid \s\tot, \d, \ldots)$, and then subsequently
sampling $\n\corr$ from its \emph{conditional} distribution with
respect to $\g$, $P(\n\corr\mid\g, \s\tot, \d, \ldots)$. These two steps must
be performed immediately after one another, or else we would introduce
an inconsistency in the Gibbs chain with respect to the other parameters.

Note that $P(\n\corr\mid\g, \s\tot, \d, \ldots)$ is unchanged compared to
the original Gibbs prescription, and no changes are required to sample
from that particular distribution \citep[see][for more details on this
sampling process]{bp06}. When it comes to ${P(\g\mid\s\tot, \d, \ldots)}$, we refer to Appendix~A.2 of \citet{bp01}, whose sampling equations we will use throughout this paper. We note that the data model used in that appendix is the same general form as Eq.~\eqref{eq:gen_data_model_scan}, and that sampling from $P(\g\mid\s\tot, \d, \ldots)$ is exactly analogous to what is shown in that appendix, as long as we make the identification $\n \rightarrow \n\corr + \n\wn$. As the covariance matrix of a sum of independent Gaussian variables ($\n\corr$ and $\n\wn$) is also Gaussian, with a covariance matrix given by the sum of the individual covariance matrices, we can in what follows use the results of the above-mentioned appendix to sample from $P(\g\mid\s\tot, \d, \ldots)$ as long as we let $\N \rightarrow \N\corr + \N\wn$.

Computationally speaking, sampling from $P(\g\mid\s\tot, \d, \ldots)$ instead of $P(\g\mid\s\tot, \n\corr, \d, \ldots)$ is numerically equivalent to a
more expensive noise covariance matrix evaluation.\footnote{Although not shown here, sampling from $P(\g\mid\s\tot, \n\corr, \d, \ldots)$ would follow the exact same procedure, but with a noise covariance matrix given by $\N\wn$ instead of $\N\wn$ + $\N\corr$. $\N\wn$ is a diagonal matrix, while $\N\corr$ is not, and since most operations are less heavy, computationally speaking, when diagonal matrices are involved, the resulting sampling process would also be lighter in that case.} To mitigate this
additional cost, we note that the gain is assumed to be slowly varying
in time, and, in particular, constant within each PID. All time-domain
operations may therefore be carried out using co-added low-resolution
data with negligible loss of precision. In practice, all TOD
operations are in the following carried out at a sample rate of 1\,Hz,
leading to a computational speed-up of about two orders of
magnitude.

\subsection{Absolute gain calibration with the orbital dipole}
Next, we also recall from Sect.~\ref{sec:gain_modelling} that the gain
is multiplicatively degenerate with the total sky signal. At the same
time, we note that the orbital CMB dipole is known to exquisite
precision, and this particular component is therefore the ideal
calibrator for CMB satellite experiments. However, its relatively low
amplitude as compared with instrumental noise renders it incapable of
tracking short-term gain variations, and, when fitted jointly with
astrophysical foregrounds, it is also not sufficiently strong to
determine relative calibration differences between detectors at the
precision required for large-scale polarization
reconstruction. Therefore, to minimize sensitivity to potential
residual systematic and modelling errors, it is advantageous to
estimate the absolute calibration using the orbital dipole alone, but
use the full signal model (including the bright Solar CMB dipole) when
estimating relative and time-dependent gain variations.

This motivates splitting the gain into two components, 
\begin{equation}
  g\qi = g_0 + \gamma\qi,
  \label{eq:split1}
\end{equation}
where $g_0$ is now independent of both time and detectors, following
\citet{planck2020-LVII}. Our goal is then to use only the orbital CMB
dipole to estimate $g_0$, and later use the full sky signal to
estimate $\gamma\qi$. Thus, with this reparametrization, we go from
sampling from $P(\g\mid\s\tot, \d, \ldots)$ to sampling from $P(g_0,
\bgamma\mid\s\tot, \d, \ldots)$. As usual, drawing samples from this
joint distribution can be done by Gibbs sampling, so that we first
sample $g_0$ from $P(g_0 \mid \bgamma, \s\tot, \d, \ldots)$ and then
$\bgamma$ from $P(\bgamma\mid g_0, \s\tot, \d, \ldots)$.

We should note that estimating $g_0$ using only the orbital dipole
formally represents a violation of the Gibbs formalism, as we no
longer draw this particular parameter from its exact conditional
distribution. This is one of many examples for which we ``buy''
stability with respect to systematic errors at the price of increased
statistical uncertainties. This is similar to the application of a
processing mask when estimating the zero-levels of a CMB sky map
\citep[e.g.,][]{planck2014-a10}, or fitting correlated noise
parameters using only a sub-range of all available temporal
frequencies \citep[e.g.,][]{bp06}. In all such cases, parts of the
data are disregarded in order to prevent potential systematic errors
from contaminating the parameter in question.
\begin{figure}[t]
  \center
  \includegraphics[width=\linewidth]{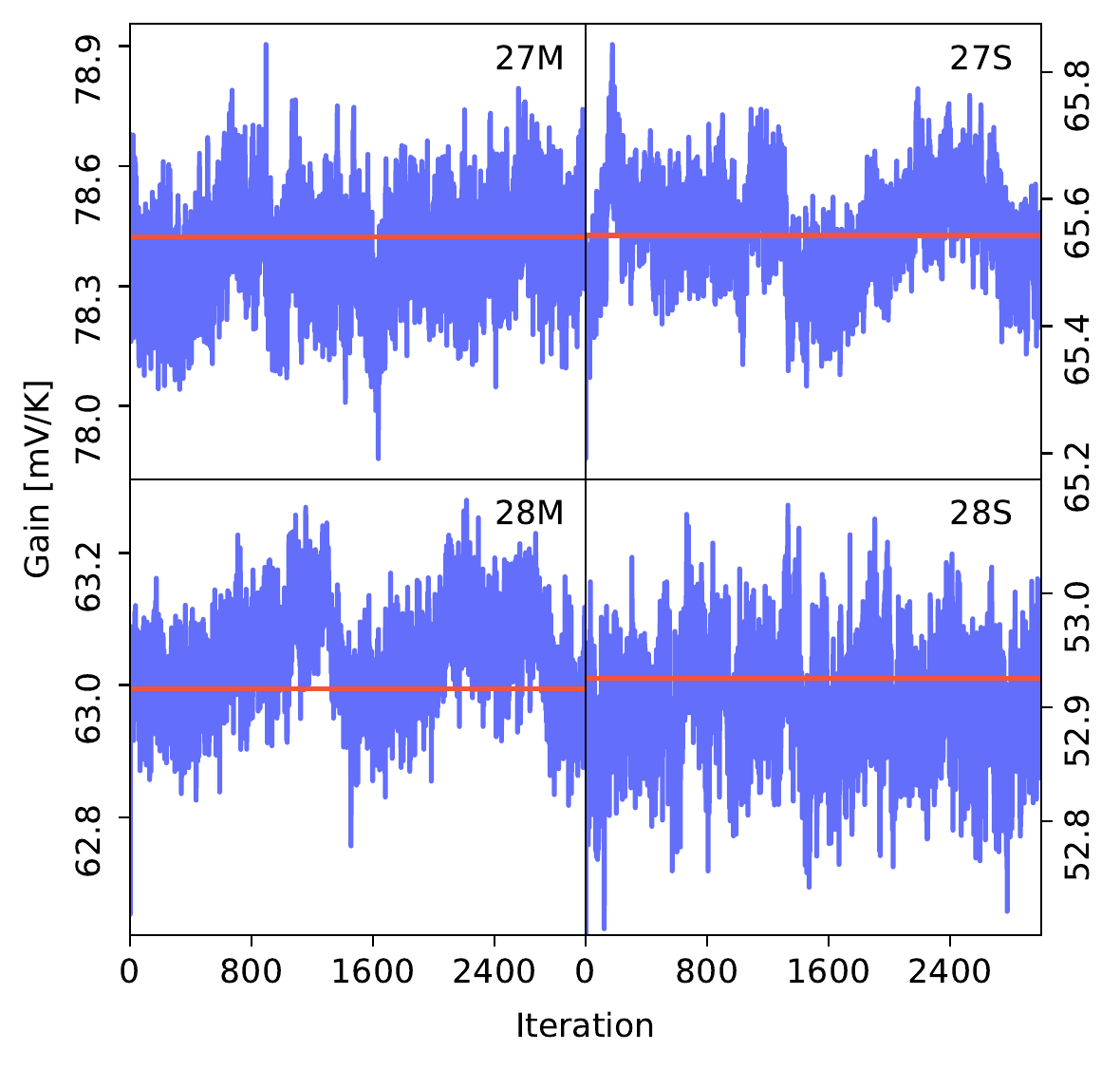}
    \caption{Trace plots of samples of the total gain for randomly selected
    PIDs for each of the four 30\,GHz detectors for our simulation run. The
    PIDs are, respectively, 349, 9847, 4298, and 1993. The red lines signify
    the input gain values.}
  \label{fig:sim_traceplots}
\end{figure}

\begin{figure}[t]
  \center
  \includegraphics[width=\linewidth]{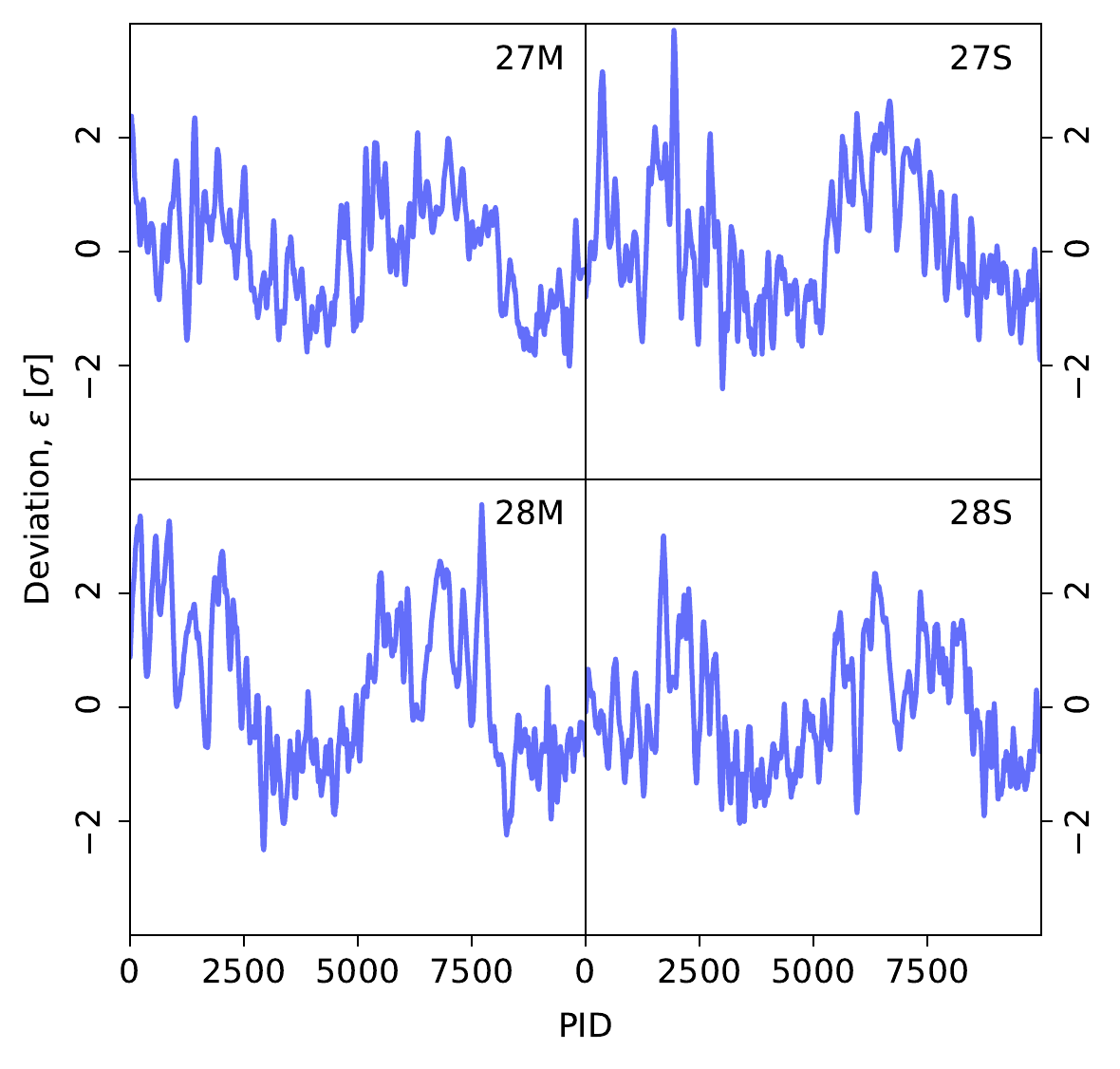}
    \caption{Deviation of the mean output gain solution from the input gain for
    each PID and 30\, GHz detector in our simulation run. The deviations are
    measured in sample standard deviations.}
  \label{fig:sim_dev}
\end{figure}
\begin{figure}[t]
  \center
  \includegraphics[width=\linewidth]{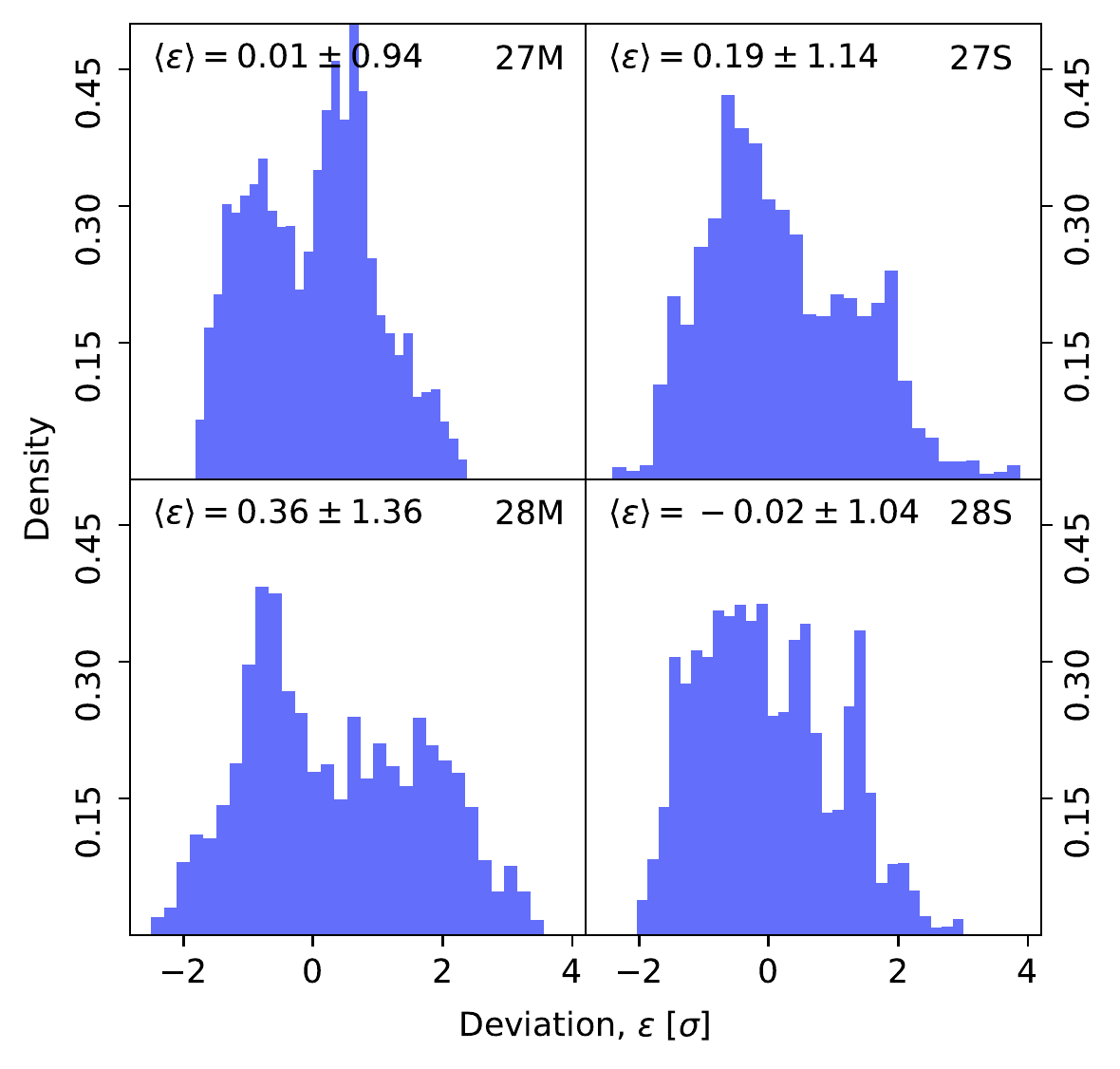}
    \caption{Histograms of the deviations of the mean output gain solution from
    the input gain, given in number of sample standard deviations in our
    simulation run. Each histogram represents the aggregation of the $10,000$
    PIDs included in the simulation validation.}
  \label{fig:sim_agghist}
\end{figure}

For the split in Eq.~\eqref{eq:split1} to be valid, we must ensure that
$\sum\qi \gamma\qi = 0$, such that $\gamma\qi$ represents only
deviations from the absolute calibration. For technical reasons, it
turns out that this will be easier to do if we also reparametrize
$\gamma\qi$,
\begin{equation}
    \gamma\qi = \Delta g_i + \delta g\qi,
\end{equation}
where $\Delta g_i$ represents the detector-specific constant gain, and
$\delta g\qi$ denotes deviations from $\Delta g_i$ per scan. We can
then separately enforce $\sum_i \Delta g_i = 0$ and $\sum_q \delta g\qi
= 0$ for each detector $i$, which is computationally cheaper than
enforcing both constraints simultaneously.

Thus, we split the gain into three nearly independent variables, and
explore their joint distribution by Gibbs sampling. The overarching
goal for this section, then, is to derive sampling algorithms for each
of the three associated conditional distributions,
\begin{align}
    & P(g_0\mid\Delta g_i, \delta g\qi, \d_i, \s_i\tot, \N_i, \ldots) \\
    & P(\Delta g_i\mid g_0, \delta g\qi, \d_i, \s_i\tot, \N_i, \ldots) \\
    & P(\delta g\qi\mid g_0, \Delta g_i, \d_i, \s_i\tot, \N_i, \ldots).
\end{align}
We now consider each of these in turn.

\subsection{Sampling the absolute calibration, $g_0$}

To sample the absolute calibration using the orbital dipole alone, we
need to define a data model that depends only on $g_0$ and $s\orb$. We
do this by first subtracting the full signal model as defined by the
current joint parameter state, and then add back only the orbital dipole term,
\begin{align}
    r\ti & \equiv d\ti -(g_0^{\mathrm{curr}} + \Delta g_i + \delta g\qi)s\tot\ti \nonumber + g_0^{\mathrm{curr}}s\orb\ti \nonumber \\
    & = g_0s\orb\ti + n\ti\tot.
    \label{eq:absolute_gain_res}
\end{align}
Here $g_0^{\mathrm{curr}}$ denotes the absolute gain at the \emph{current}
iteration in the Gibbs chain, i.e., before drawing a new value for
$g_0$. 

As noted earlier, working with this residual and using the previous
sample of $g_0$ to estimate the current sample does represent a
breaking of the Gibbs formalism, since the statistically exact
residual for $g_0$ would be
\begin{equation}
    d\ti - (\Delta g_i + \delta g\qi)s\tot\ti = g_0 s\tot\ti + n\ti\tot.
\end{equation}
However, in this case we would also be calibrating $g_0$ on the total
sky signal instead of just the orbital dipole. Thus, we trade
mathematical rigour and statistical uncertainties for stronger
robustness with respect to systematic effects.

\begin{figure}
  \center
  \includegraphics[width=\linewidth]{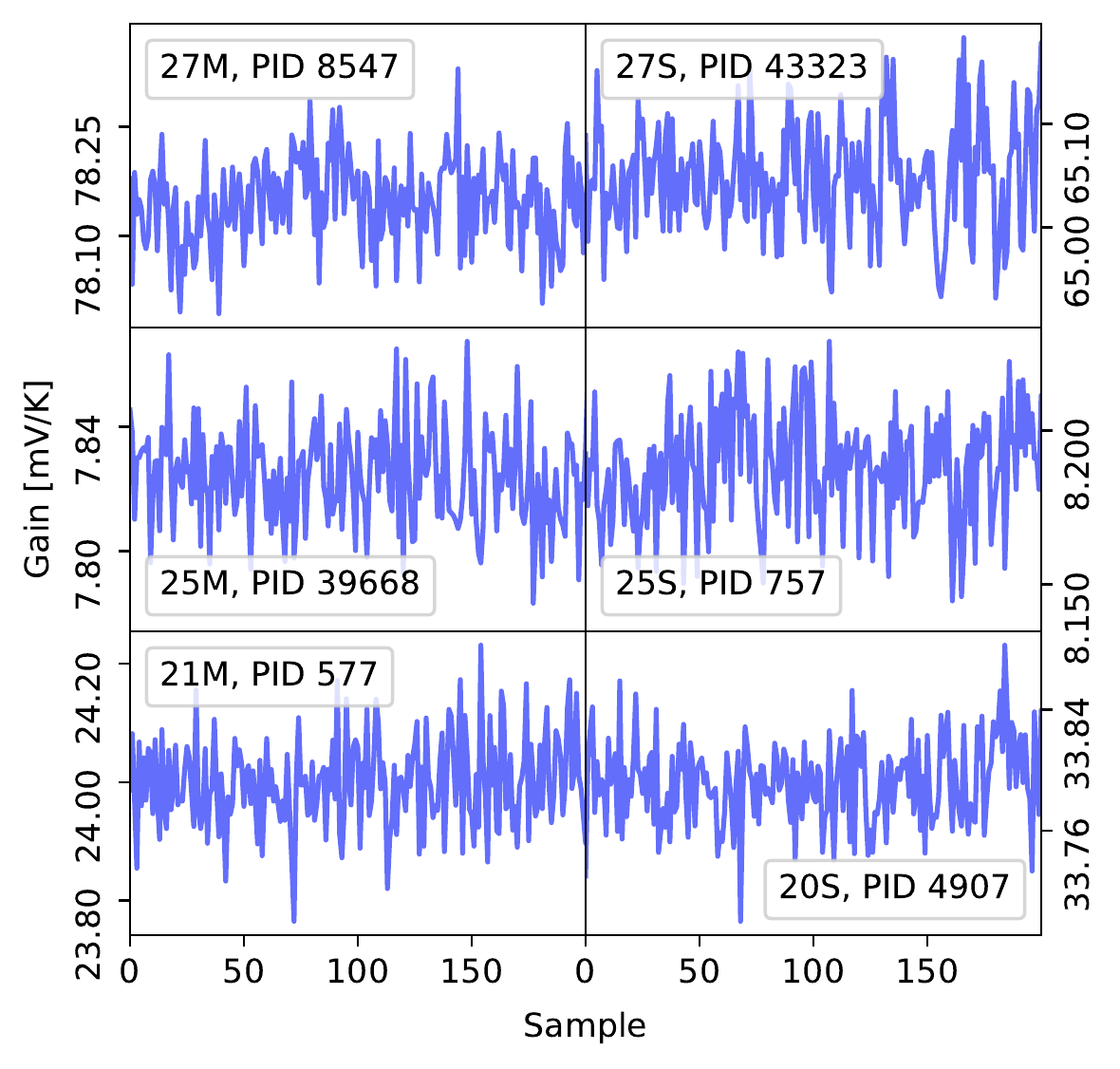}
    \caption{Gibbs chains of the total gain for selected detectors and PIDs.}
  \label{fig:chains}
\end{figure}

\begin{figure}[t]
  \center
    \includegraphics[width=0.8\linewidth]{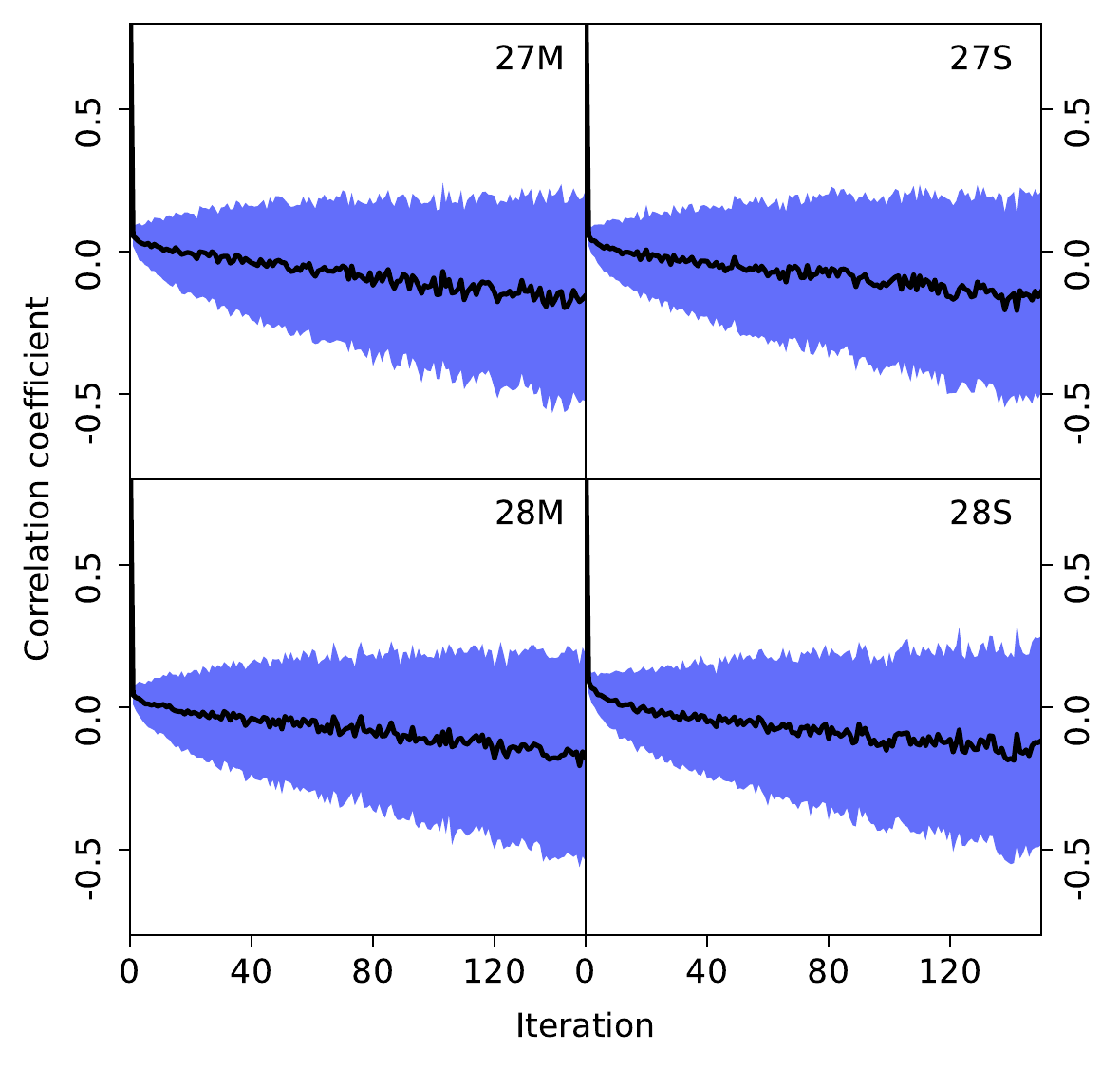}\\\vspace*{-2mm}
    \includegraphics[width=0.8\linewidth]{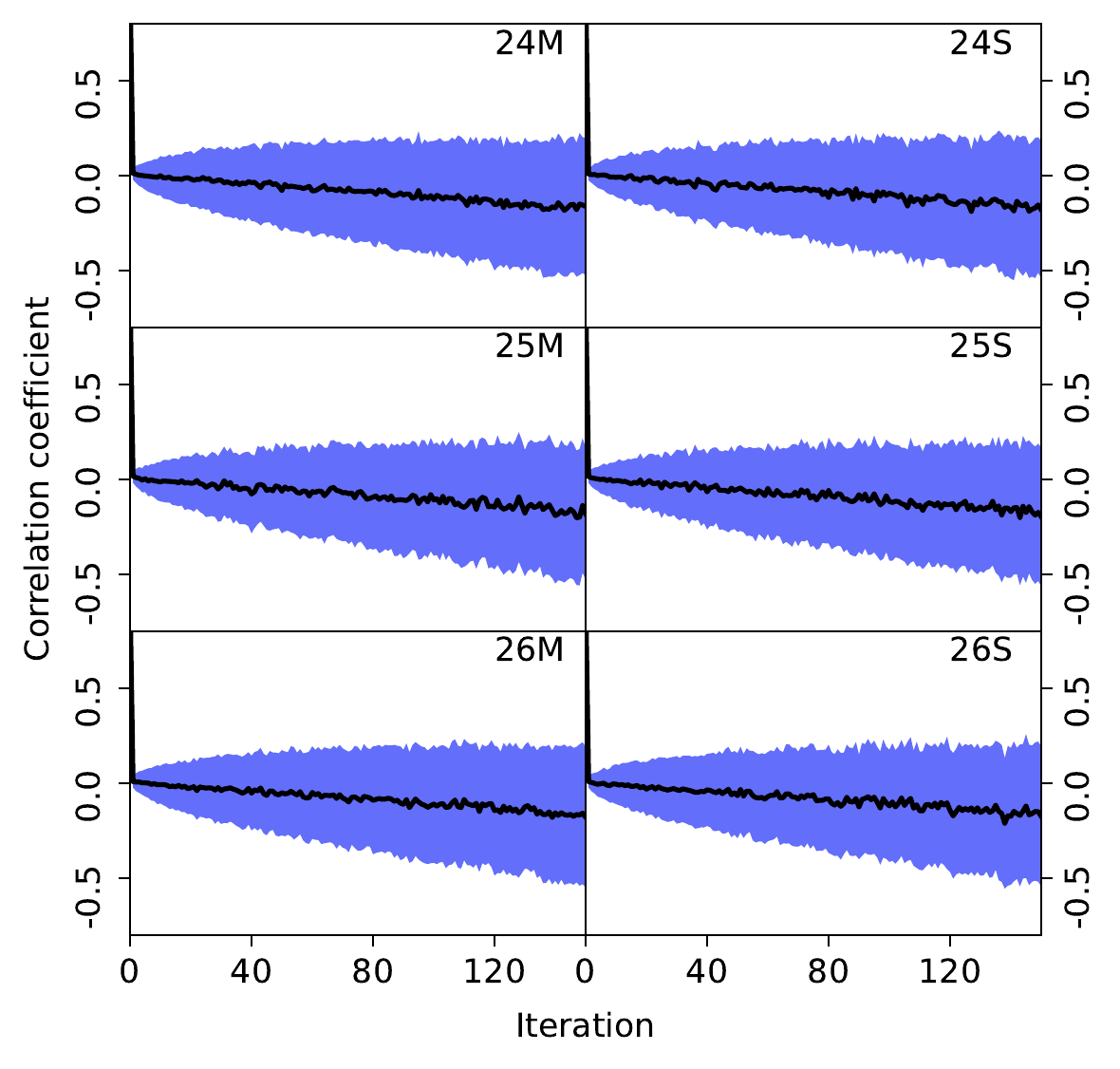}\\\vspace*{-2mm}
    \includegraphics[width=0.8\linewidth]{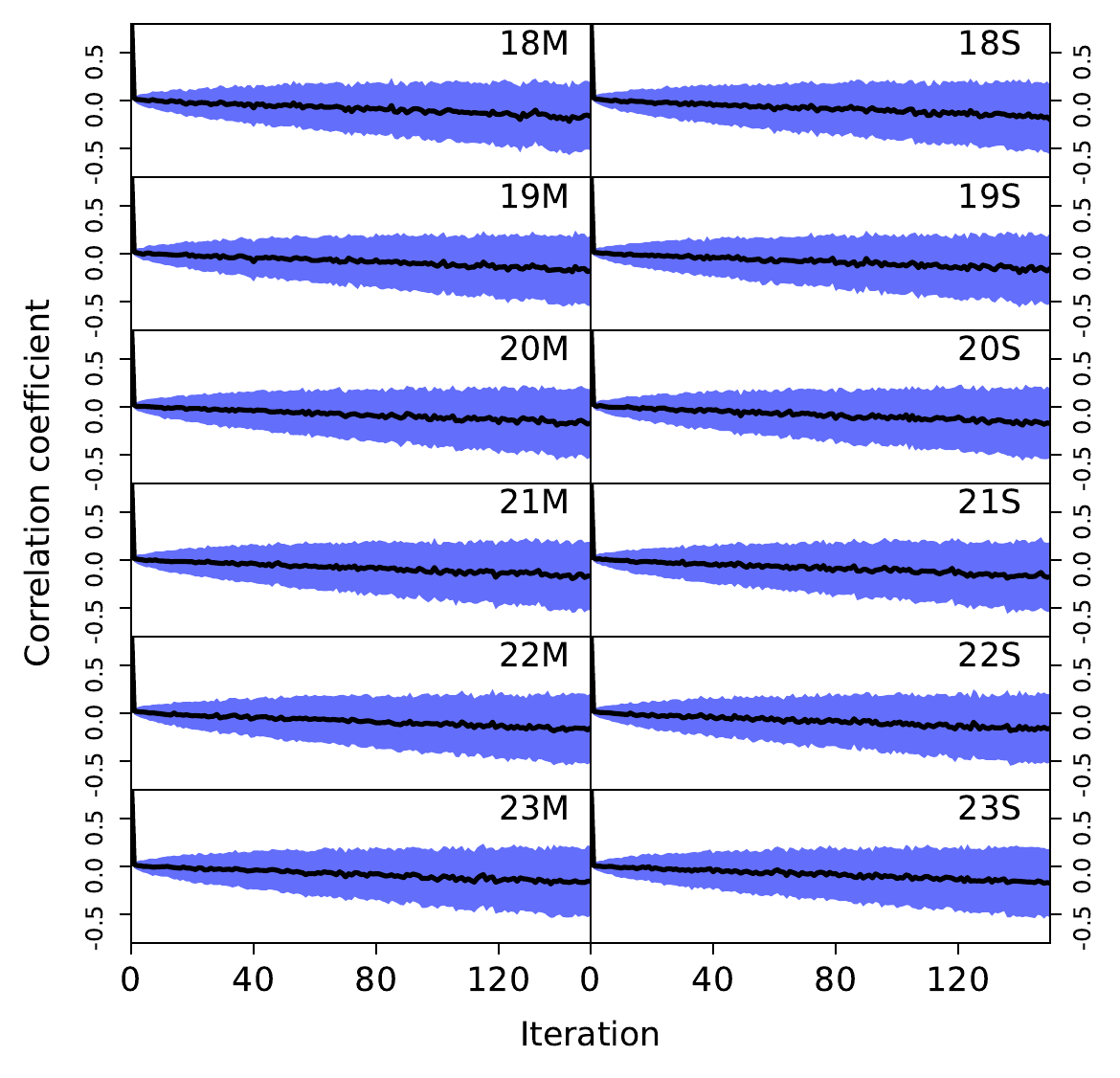}

    \caption{Correlation coefficients as a function of distance between Gibbs samples for 30 (\emph{top panel}), 44 (\emph{middle panel}), and 70\,GHz (\emph{bottom panel}) detectors. The black thick line shows the mean value for all PIDs, while the blue band shows the $1\sigma$ error bars.}
  \label{fig:corrlengths}
\end{figure}
As discussed in Sect.~\ref{sec:corrnoise}, the noise term in
Eq.~\eqref{eq:absolute_gain_res} includes both correlated and white
noise, and the appropriate covariance matrix is therefore $\N =
\N\corr + \N\wn$.  Given this fact, Eq.~\eqref{eq:absolute_gain_res}
corresponds to a simple uni-variate Gaussian model as a function of
$g_0$, and the appropriate sampling algorithm is discussed in
Appendix~A.2 of \citet{bp01}.  Applying that general procedure to our
special case, we may write down the following sampling equation for
$\hat{g}_0$,\footnote{Note that we do not apply any priors on $g_0$ in
  this paper, which corresponds to $\S^{-1}=0$, adopting the notation
  of \citet{bp01}, where $\S$ is the prior covariance of $g_0$. The
  remaining notational differences between
  Eq.~\eqref{eq:absolute_gain_sampling} and Eq.~(A.10) in that paper
  arise from our organizing all vectors and matrices in terms of
  independent detectors, using the fact that $\n\tot$ is assumed to be
  independent between detectors; this may not be strictly true in
  practice, as discussed by \citet{bp06}, and future analyses may
  prefer to account for the full joint matrix.}
\begin{equation}
    \hat{g}_0 = \frac{\sum_i (\vec{s}_i^{\mathrm{orb}})^T\N^{-1}_i\vec{r}_i}{\sum_i (\vec{s}_i^{\mathrm{orb}})^T\N_i^{-1}\vec{s}_i^{\mathrm{orb}}} + \frac{\eta}{\sqrt{\sum_i(\vec{s}_i^{\mathrm{orb}})^T\N_i^{-1}\vec{s}_i^{\mathrm{orb}}}},
    \label{eq:absolute_gain_sampling}
\end{equation}
where $\eta\sim N(0,1)$ is a random number drawn from a standard
normal distribution. Here, and elsewhere, a $T$ superscript indicates
the matrix transpose operator. The first term in this equation is the (Wiener
filter) mean of the distribution $P(g_0\mid\r_i, \N_i)$, while the second
term ensures that $g_0$ has the correct variance.

\subsection{Sampling detector-dependent calibration}
\label{sec:delta_gi_sampling}
For $\Delta g_i$, we proceed similarly as with $g_0$, with two
exceptions. First, $\Delta g_i$ now represents the relative
calibration between detectors, and, as discussed in
Sect.~\ref{sec:gain_modelling}, we need to use a stronger calibration
signal than the orbital dipole to avoid significant polarization
leakage. Secondly, we have to impose the constraint $\sum_i \Delta g_i
= 0$.

We start by defining the following residual,
\begin{equation}
    r_{t,i} \equiv d_{t,i} - (g_0 + \delta{g}_{q,i})s_t^{\mathrm{tot}} = \Delta g_i s_t^{\mathrm{tot}} + n_t^{\mathrm{tot}}
    \label{eq:relative_gain_res}
\end{equation}
for each detector. This equation is structurally similar to
Eq.~\eqref{eq:absolute_gain_res}, with the main difference being that
the total sky signal, which is dominated by the Solar dipole, is
retained on the right-hand side. Otherwise,
Eq.~\eqref{eq:relative_gain_res} still represents a Gaussian model,
and we should be able to proceed similarly as for $g_0$ when drawing
from the conditional distribution. We do, however, need to ensure that
$\sum_i \Delta g_i=0$, and this will significantly impact the form of
the target distribution. The numerically most convenient method for
imposing such a constraint is through the method of Lagrange
multipliers.

In general, the method of Lagrange multipliers allows the user to
minimize a function $f(\x)$ under some set of constraints that may be
formulated as $g(\x) = 0$. Without these constraints, one would of
course determine $x$ by solving $df/dx=0$. With additional
external constraints, however, it is convenient to instead define the
so-called Lagrangian,
\begin{equation}
    \mathcal{L}(\x, \lambda) = f(\x) + \lambda g(\x),
    \label{eq:lagrangian}
\end{equation}
and set the corresponding partial derivatives with
respect to $\x$ and $\lambda$ equal to zero. It is readily seen that
$\partial\mathcal{L}/\partial \lambda = 0$ corresponds
directly to $g(\x) = 0$, which is precisely the desired
constraint.

Our primary target distribution is
\begin{align}
    P(\Delta g \mid \r, \s\tot, \N) &\propto P(\r \mid\Delta g, \s\tot, \N)P(\Delta g) \nonumber \\
    &\propto \exp\left(\sum_i\left(\r_i - \Delta g_i \s^{\mathrm{tot}}\right)^T\N_i^{-1}\left(\r_i - \Delta g_i\s^{\mathrm{tot}}\right)\right)
\end{align}
where the first line follows from Bayes' theorem, and the second
follows from the fact that we assume vanishing covariance between
detectors, and that $\r_i$ is Gaussian distributed with a mean of
$\Delta g_i \s\tot_i$ and covariance $\N_i$. We are of course free to minimize
the logarithm of this function instead of the function itself, which
makes things easier as it takes the exponential away. We may
therefore define the following Lagrangian,
\begin{equation}
    \mathcal{L}(\Delta g_i, \lambda) = \sum_i\bigl(\r_i - \Delta g_i \s^{\mathrm{tot}}\bigr)^T\N_i^{-1}\bigl(\r_i - \Delta g_i\s^{\mathrm{tot}}\bigr) + \lambda\sum_i \Delta g_i,
\end{equation}
where $\lambda$ is the Lagrange multiplier. 

To optimize this function, we take the derivative with respect to
$\Delta g_i$ and $\lambda$ to obtain two coupled equations. The first
equation takes the form
\begin{align}
    \frac{\partial \mathcal{L}}{\partial \Delta g_i}  &= 0 \nonumber \\
    \Rightarrow -2\bigl( \r_i - \Delta g_i\s_i^{\mathrm{tot}}\bigr)^T\N^{-1}_i\s_i^{\mathrm{tot}} + \lambda &= 0 \nonumber \\
    \Rightarrow \Delta g_i (\s_i^{\mathrm{tot}})^T\N^{-1}_i\s_i^{\mathrm{tot}} + \frac{1}{2}\lambda &= (\r_i)^T \N^{-1}_i s_i^{\mathrm{tot}},
    \label{eq:lagrange_eq_1}
\end{align}
while the second simply reads
\begin{align}
    \frac{\partial \mathcal{L}}{\partial \lambda} & = 0 \nonumber \\
    \Rightarrow \sum_i \Delta g_i & = 0.
\end{align}

Jointly solving these linear equations for $\Delta g_i$ provides
estimates with the correct mean. What we require, however, is a sample
from the appropriate distribution, and not mean estimates. We must
therefore add a fluctuation term, as in
Eq.~\eqref{eq:absolute_gain_sampling}. To do so, we note that if it
were not for $\lambda$, Eq.~\eqref{eq:lagrange_eq_1} would have the
exact same form as Eq.~\eqref{eq:absolute_gain_sampling}, with $\s\tot$ substituted for $\s\orb$. Comparing
Eq.~\eqref{eq:lagrange_eq_1} with Eq.~\eqref{eq:absolute_gain_sampling},
we then see that the final equation for the desired sample must be
\begin{equation}
	\Delta\hat{g}_i (\s_i\tot)^T\N^{-1}_i\s_i\tot + \frac{1}{2}\lambda = (\r_i)^T\N_i^{-1}\s_i\tot + \eta\sqrt{(\s\tot_i)^T\N_i^{-1}\s\tot_i},
\end{equation}
where, as usual, $\eta\sim N(0,1)$.

Casting this in terms of a linear system with $n_{\mathrm{detector}}
+1$ unknowns, this may be solved straightforwardly with standard
numerical libraries. For a two-detector example, the resulting system
of equations takes the form
\begin{align}
    \begin{bmatrix}
        (\s^{\mathrm{tot}})^T \N_1^{-1} \s^{\mathrm{tot}} & 0 & \frac{1}{2} \\
        0 &(\s^{\mathrm{tot}})^T \N_2^{-1} \s^{\mathrm{tot}} & \frac{1}{2} \\
        1 & 1 & 0
    \end{bmatrix}
    \begin{bmatrix}
        \Delta \hat{g}_1 \\
        \Delta \hat{g}_2 \\
        \lambda
    \end{bmatrix}
    = \\ 
    \begin{bmatrix}
        (\r_1)^{\mathrm{tot}} \N_1^{-1}\s_1^{\mathrm{tot}} + \eta_1\sqrt{(\s_1\tot)^T\N_1^{-1}\s_1\tot}\\
        (\r_2)^{\mathrm{tot}} \N_2^{-1}\s_2^{\mathrm{tot}} + \eta_2\sqrt{(\s_2\tot)^T\N_2^{-1}\s_2\tot}\\
        0
    \end{bmatrix}.
    \label{eq:relative_gain_sampling}
\end{align}

\subsection{Sampling temporal gain variations with Wiener filter smoothing}
\label{sec:temporal_var_sampling}
Finally, we consider the temporal gain variations, $\delta g\qi$. As
before, we write down the following residual,
\begin{equation}
    r_{t,i}\equiv d_{t,i} - (g_0 + \Delta g_i)s\ti\tot = \delta g\qi s\ti\tot + n\ti\tot,
    \label{eq:delta_gain_residual}
\end{equation}
where we again employ the total signal as a calibrator. The only
difference with respect to Eq.~\eqref{eq:relative_gain_res} is that
$\delta g\qi$ now contains multiple elements per detector, and is now
a vector in PID space. We can make this point more explicit by writing
\begin{equation}
    \r_i \equiv \d_i - (g_0 + \Delta g_i)\s_i\tot = \T_i\delta \g_i + \n_i\tot,
\end{equation}
where $\T$ is an $n_{\mathrm{samp}}\times n_{\mathrm{scan}}$-matrix
that contains $s\ti\tot$ in element $(t,q)$ for all values of $t$ in
scan $q$. All other elements are zero. Thus, $\T$ projects $\delta \g_i$
into the $n_{\mathrm{samp}}$-dimensional space of $\r_i$ and
$\n_i\tot$.

Once again following the procedure in Equation~A.3 in \citet{bp01}, we
may write down the following sampling equation,
\begin{equation}
    \T_i^T\N_i^{-1}\T_i\,\delta \hat{\g}_i = \T_i^T\N_i^{-1}\r_i + (\T_i^T\N_i^{-1}\T_i)^{\frac{1}{2}}\vec\eta,
    \label{eq:delta_gain_sampling}
\end{equation}
where $\vec\eta\sim N(\vec 0,\tens I)$ is a random Gaussian vector of length
$n_{\mathrm{scan}}$.

In its current form, Eq.~\eqref{eq:delta_gain_sampling} assumes that
$\delta g\qi$ is uncorrelated between scans. As discussed by
\citet{planck2013-p03f}, \citet{planck2014-a09}, \citet{planck2016-l03}, and \citet{planck2020-LVII}, this is
not an efficient model for the \Planck\ LFI data, because the gain is
primarily determined by the thermal environment of the instrument, which is
quite stable in time. It is therefore advantageous, and in practice necessary,
to enforce some form of smoothing between $\delta g\qi$ to obtain robust
results.

\begin{figure}
  \center
  \includegraphics[width=\linewidth]{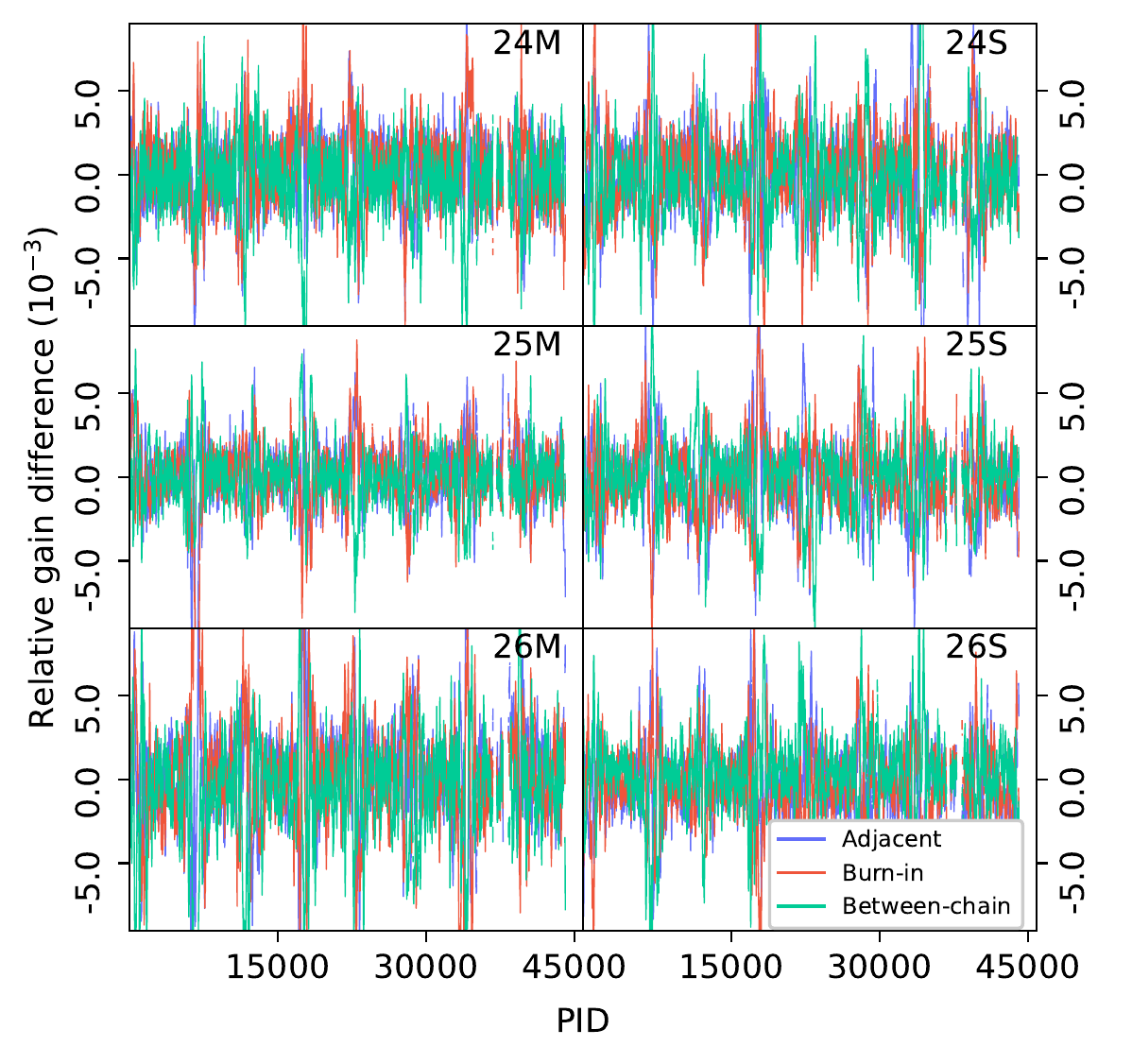}
    \caption{Relative differences between the last sample in one of the \BP\ chains and the sample adjacent to the last sample (blue line), the initial sample (the burn-in difference; red line) and the last sample in another chain (green line), for selected detectors.}
  \label{fig:allpid_gains}
\end{figure}

\begin{figure*}[t]
  \center
  \includegraphics[width=\linewidth]{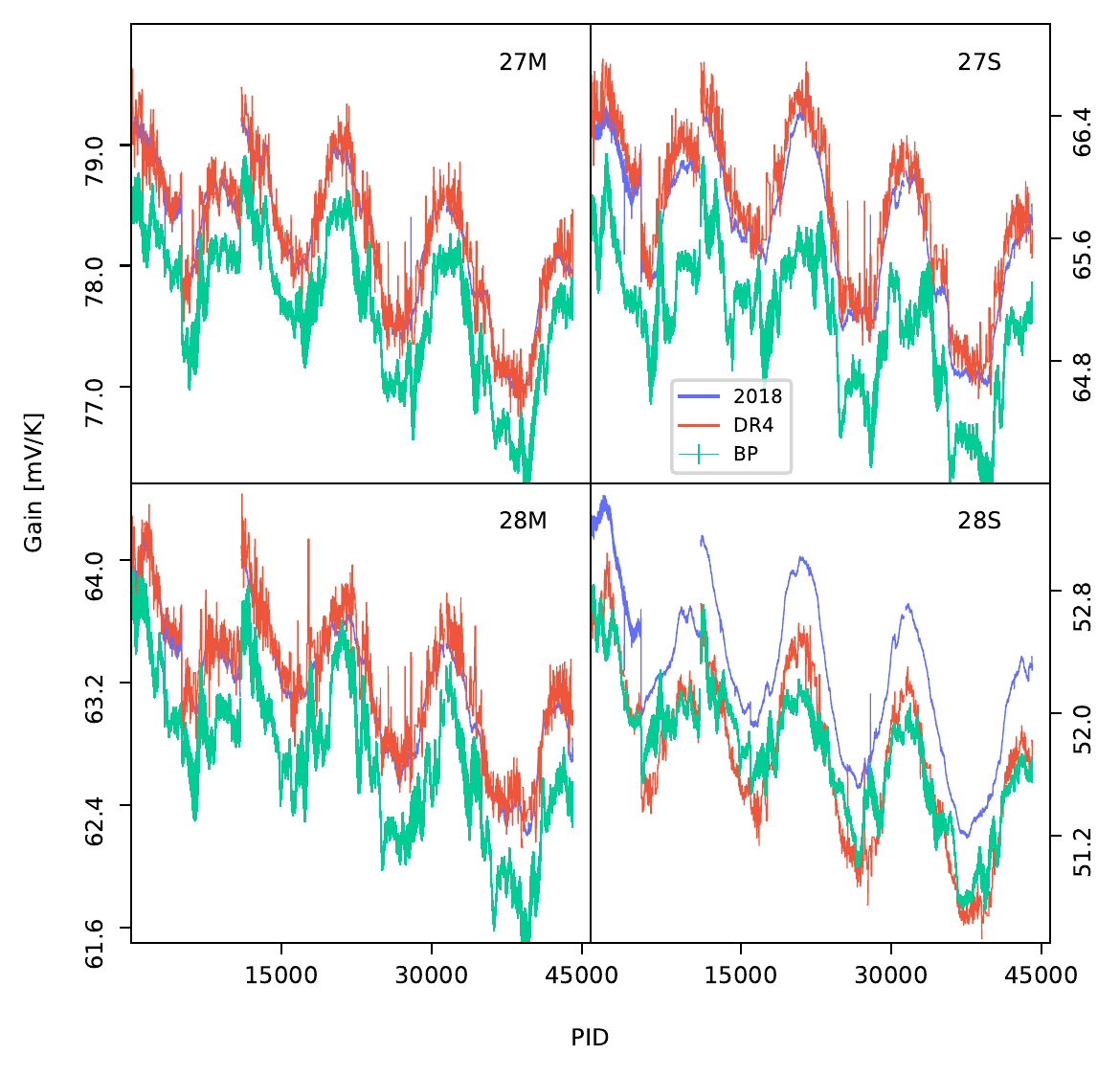}
    \caption{Comparison of gain estimates for the 30\,GHz detectors for \Planck\ 2018, \Planck\ DR4, and \BP. 
    The width of the \BP\ line is given by the Monte Carlo uncertainty of the chains.}
  \label{fig:gaincomp_030}
\end{figure*}

\subsubsection{Wiener filtering}
\begin{figure*}[t]
  \center
  \includegraphics[width=\linewidth]{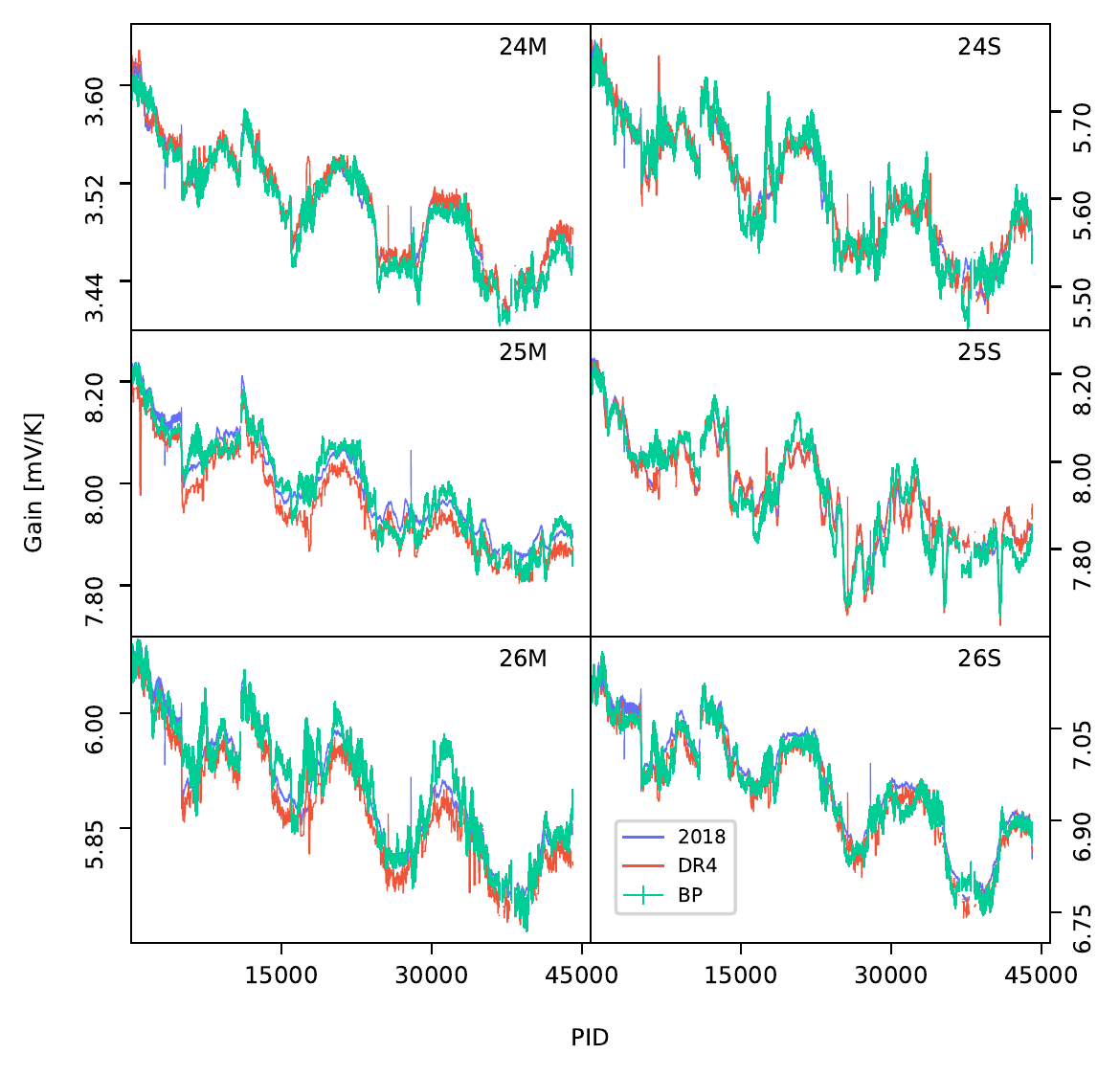}
    \caption{Comparison of gain estimates for the 44\,GHz detectors. 
    See Fig.~\ref{fig:gaincomp_030} for details.}
  \label{fig:gaincomp_044}
\end{figure*}

The smoothing approach we adopt here is \emph{Wiener filtering}. Although this approach has been applied to other parts of the CMB analysis pipeline such as CMB and noise estimation, it has never before been applied to the gain estimation process. In Bayesian terms, we have so far been drawing samples from the \emph{likelihood function} $\mathcal{L}(\delta \g_i)$, which, since our actual goal is to draw samples from the \emph{posterior} distribution of $\delta \g_i$, means that we are implicitly assuming a uniform prior on this parameter, which in turn means that we let the estimates of $\delta \g_i$ be completely determined by the data alone. 

In addition, this means we assume no correlations between the elements in the $\delta \g_i$ vector, as the likelihood function is separable into independent probability distribution functions -- an effect of this is that Eq. \eqref{eq:delta_gain_sampling} is really a set of independent equations that can be solved sequentially. 
\begin{figure*}[t]
  \center
  \includegraphics[width=\linewidth]{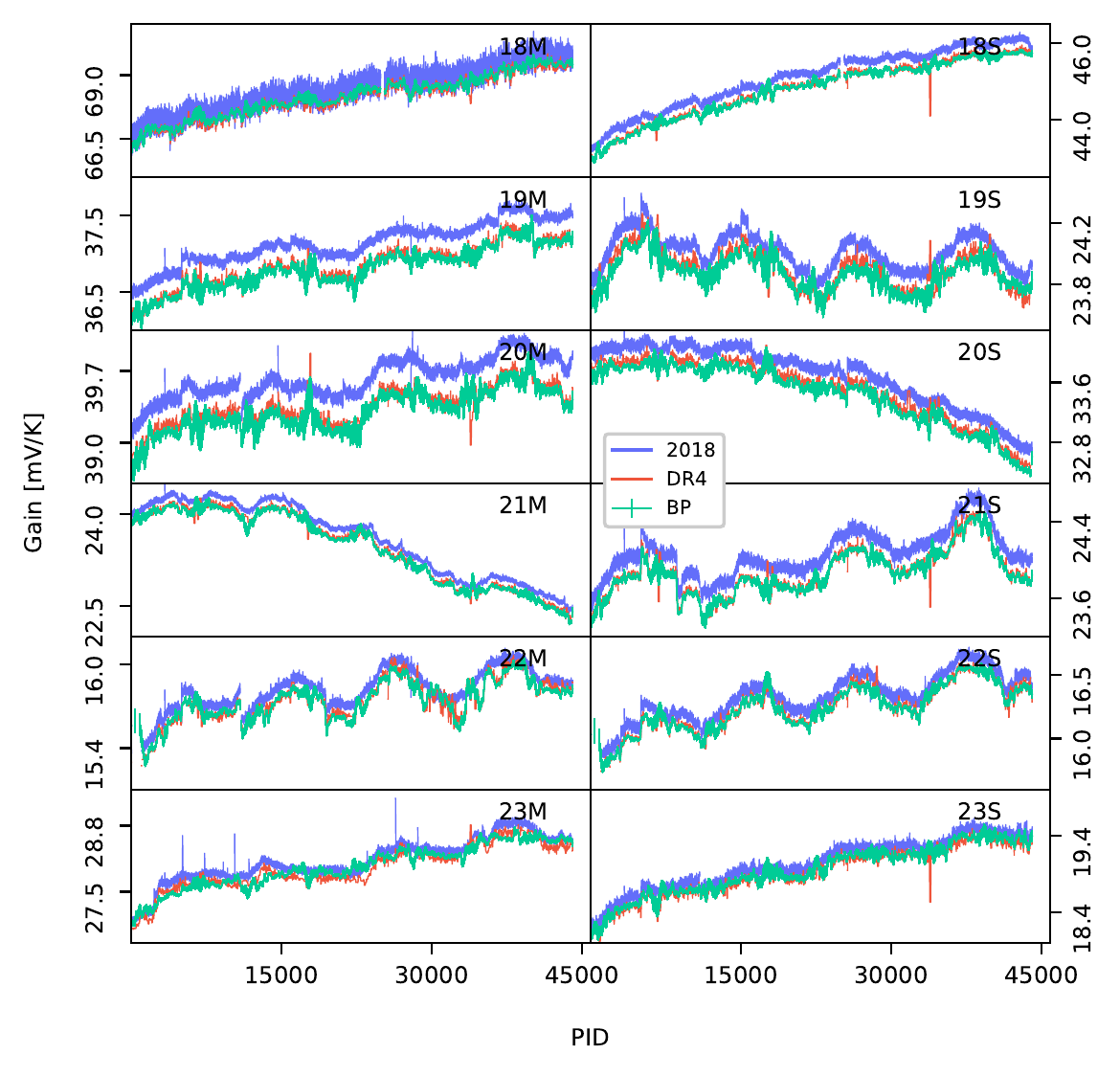}
    \caption{Comparison of gain estimates for the 70\,GHz detectors. See Fig.~\ref{fig:gaincomp_030} for details.}
  \label{fig:gaincomp_070}
\end{figure*}

Wiener filtering the data essentially amounts to applying a Gaussian prior to the estimation process, meaning that instead of drawing samples from $\mathcal{L}(\delta \g_i)$ alone, we draw samples from $\mathcal{L}(\delta \g_i) P(\delta \g_i)$, where $P(\delta \g_i)$ is the Gaussian prior we want to apply, which in turn will depend on how we \emph{a priori} expect the gain fluctuations to behave. This process will by construction ensure that for scans with high signal-to-noise, the gain estimates will be set mainly by the observed data, whereas in periods of lower signal-to-noise (such as for dipole minima) the estimates will be prior dominated (and thus less fluctuating than without any prior). In addition, this prior explicitly introduces (physically motivated) correlations between the gain fluctuations of different scans which are taken into account during the sampling process rather than being artificially applied after the fact.

In order to sample from the posterior, we need to draw a sample from the product of two Gaussians, $\mathcal{L}(\delta g\qi)$ and $P(\delta g\qi)$:

\begin{equation}
P(\delta \g_i \mid \d_i) \propto \mathcal{L}(\delta \g_i)P(\delta \g_i).
\end{equation}

The proper way to sample from such a distribution is given by Eq.~(A.10) in \citet{bp01}. In this case, the procedure amounts to replacing Eq. \eqref{eq:delta_gain_sampling} by 
\begin{equation}
    (\tG^{-1} + \T_i^T\N_i^{-1}\T_i)\delta \hat{\g}_i = \T_i^T\N_i^{-1}\r_i + \T_i^T\N_i^{-1/2}\vec\eta_1 + \tG^{-1/2}\vec\eta_2,
    \label{eq:delta_gain_sampling_with_prior}
\end{equation}
where $\tG$ is the covariance matrix of the Gaussian prior on $\delta\g\qi$, and where $\eta_1$ and $\eta_2$ are two independent vectors drawn from a normal distribution with unity variance. 

Eq.~\eqref{eq:delta_gain_sampling_with_prior} is, in contrast to Eq.~\eqref{eq:delta_gain_sampling}, no longer possible to solve scan-by-scan, as the covariance matrix $\tG$ introduces cross-scan correlations. To invert this system, we use the conjugate gradient solver procedure described in \citet{bp01}.

Our prior should reflect our knowledge about the gain fluctuations - namely, that such fluctuations are related to fluctuations in the detector electronics. Such fluctuations are generally modelled as having a so-called "pink noise" or $1/f$ spectrum, which means that the gain fluctuation covariance matrix $\tG$ can be written as
\begin{equation}
    \tG(f) = \sigma_0^2\left(\frac{f}{f_{\textrm{knee}}}\right)^\alpha.
\end{equation}
Here, $\alpha$, $f_{\textrm{knee}}$ and $\sigma_0$ are parameters to be determined; for a complete marginalization over these degrees of freedom, they can be made part of the general Gibbs chain in which we are sampling the gains, and an additional sampling step for these parameters can be inserted after sampling the gain fluctuations. This is similar to the process described in \citet{bp06}, but we only need to sample a single set of parameters per detector, not one set per scan as in the correlated noise case, since the data points in the gain fluctuation case are the scan themselves, not the individual detector samples.

In the case of LFI processing, we have found that fixing $\alpha$, $\sigma_0$ and $f_{\textrm{knee}}$ instead of sampling over them gives the seemingly best results in terms of gain stability and CMB solutions. For the current paper, through trial, error, and by-eye inspection we've found that fixing the parameters to the following values makes the prior behave as desired: $\alpha = -2.5$, $\sigma_0 = 3\cdot 10^{-4}\,\textrm{V}^2/\textrm{K}^2$, $f_{\textrm{knee}} = 1\, \textrm{hr}^{-1}$. Although this might seem like an arbitrary choice, the fine-tuning represented here is not substantially different from the fine-tuning of window widths needed for the boxcar smoothing approach used by the \Planck\ DPC.

\section{Validation by controlled simulations}

As described by \citet{bp04}, the \BP\ analysis framework includes a
simulation tool that allows us to input a controlled simulation whose
aspects are all perfectly known, and we can use this to validate the
reconstructed posterior distributions. The main results from a
simulation that considers only one year of LFI 30\,GHz observations
are presented by \citet{bp04}. Here, we reproduce some of those
results that pertain to the gain estimation results. This simulation
includes only CMB (fluctuations, Solar and orbital dipole), correlated
and white noise, and gain fluctuations. We then perform an end-to-end
\commander\ analysis in which we sample $\a^{\mathrm{CMB}}$,
$\n^{\mathrm{corr}}$, and $\g$, with no other ancillary data; this is
thus a test of the core gain estimation, correlated noise estimation,
and mapmaking routines, but not of, say, component separation or
sidelobe estimation; those are validated through separate means. The
analysis comprises 3,000 samples and the first 1,000 are discarded
as burn-in (although these are still included in the trace-plots
below).

First, we show in Fig.~\ref{fig:sim_traceplots} the total gain samples
as a function of Gibbs iteration for one randomly chosen PID for each
of the four 30\,GHz detectors, and the horizontal line shows the true
input. These trace-plots show both that the chains fluctuate around the
input gain values, and that their scatter provides a meaningful
estimate of the uncertainties. However, we also see that the
correlation lengths are significant. This is due to a strong
degeneracy between the absolute gain and the CMB Solar dipole when
analyzing only a small subset of the data, in this case only one year
of 30\,GHz observations. When analyzing jointly all data from all
channels, the CMB Solar dipole signal-to-noise ratio increases
dramatically, and the correlation length goes down, as shown in Sec.~\ref{sec:results}.

To validate all scans and detectors, Fig.~\ref{fig:sim_dev} shows the
relative reconstruction bias measured in units of standard deviations, i.e.,
\begin{equation}
  \epsilon \equiv \frac{g_{q, i}^{\mathrm{in}} - \overline{g}_{q, i}}{ \sigma_{q, i}}
\end{equation}
where $g_{q, i}^{\mathrm{in}}$ is the true input total gain, $\overline{g}_{q, i}$
is the mean sample total gain estimated over all the samples for a
given PID and detector, and $\sigma_{q, i}$ is the sample standard
deviation. Thus, the figure shows the deviation of the output gain
solution from the input in units of standard deviation. Overall, most
samples lies within $\pm2\,\sigma$, although there are a few notable
outliers. It is also worth noting that the gains are intrinsically
correlated in time, and this causes the apparent correlation in this
figure. 

Finally, in Fig.~\ref{fig:sim_agghist} we show corresponding
histograms of the same quantity, but this time aggregated over all
PIDs for a given detector. Ideally, these should be Gaussian
distributed with zero mean and unit standard deviation. Overall, we
see that gain values are generally well-recovered with small biases
and reasonable uncertainties. The non-Gaussian features are due to the
long correlation lengths seen in Fig.~\ref{fig:sim_dev}, which implies
that the number of independent samples in these functions is limited.
\begin{figure}[t]
    \center
    \includegraphics[width=\linewidth]{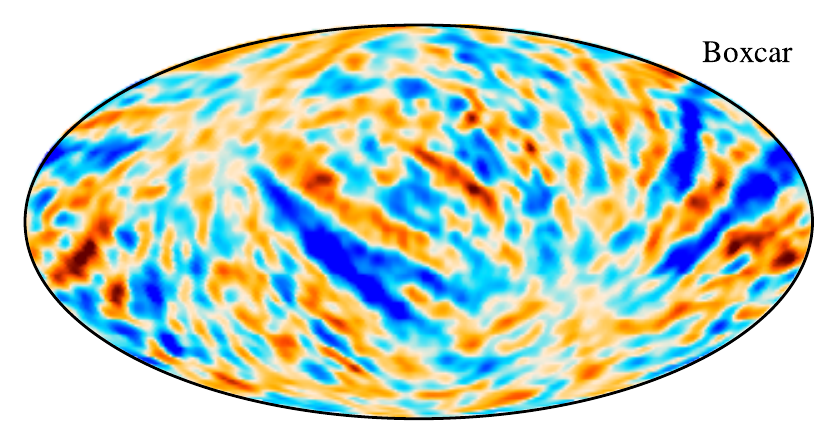}\\\vspace*{-3mm}
    \includegraphics[width=\linewidth]{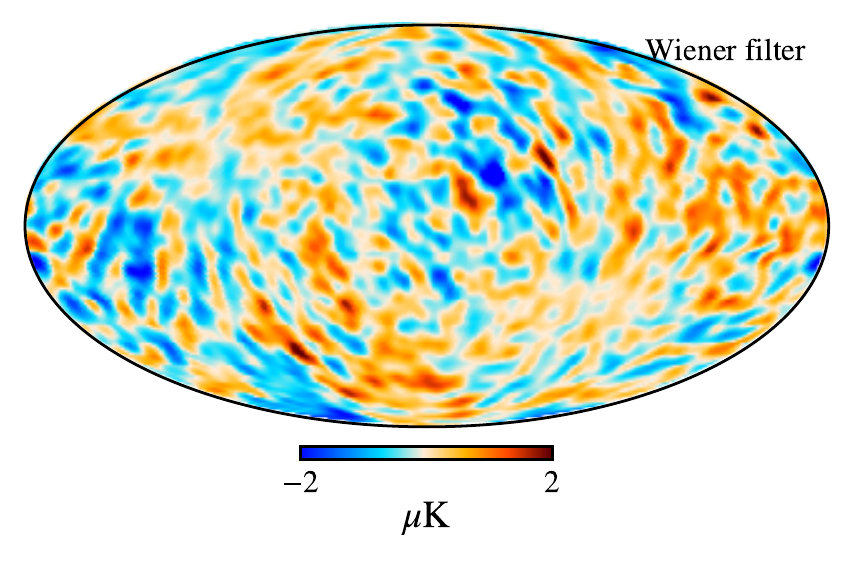}
    \caption{Map of the correlated noise, $\n^{\mathrm{corr}}$, of the $Q$ Stokes parameter for the 44\,GHz
    frequency channel, smoothed to an effective angular resolution of
    $5^{\circ}$ FWHM. The top figure is the map resulting from a
    boxcar smoothed gain solution, whereas the bottom figure is the map which
    results from smoothing the gain solution with a Wiener filter.
    }\label{fig:corrstripes}
\end{figure}

\section{Results}
\label{sec:results}

We are now finally ready to present gain estimates for each
\Planck\ LFI radiometer, as estimated within the end-to-end Bayesian
\BP\ analysis framework. 
\begin{figure}[t]
  \center
    \includegraphics[width=\linewidth]{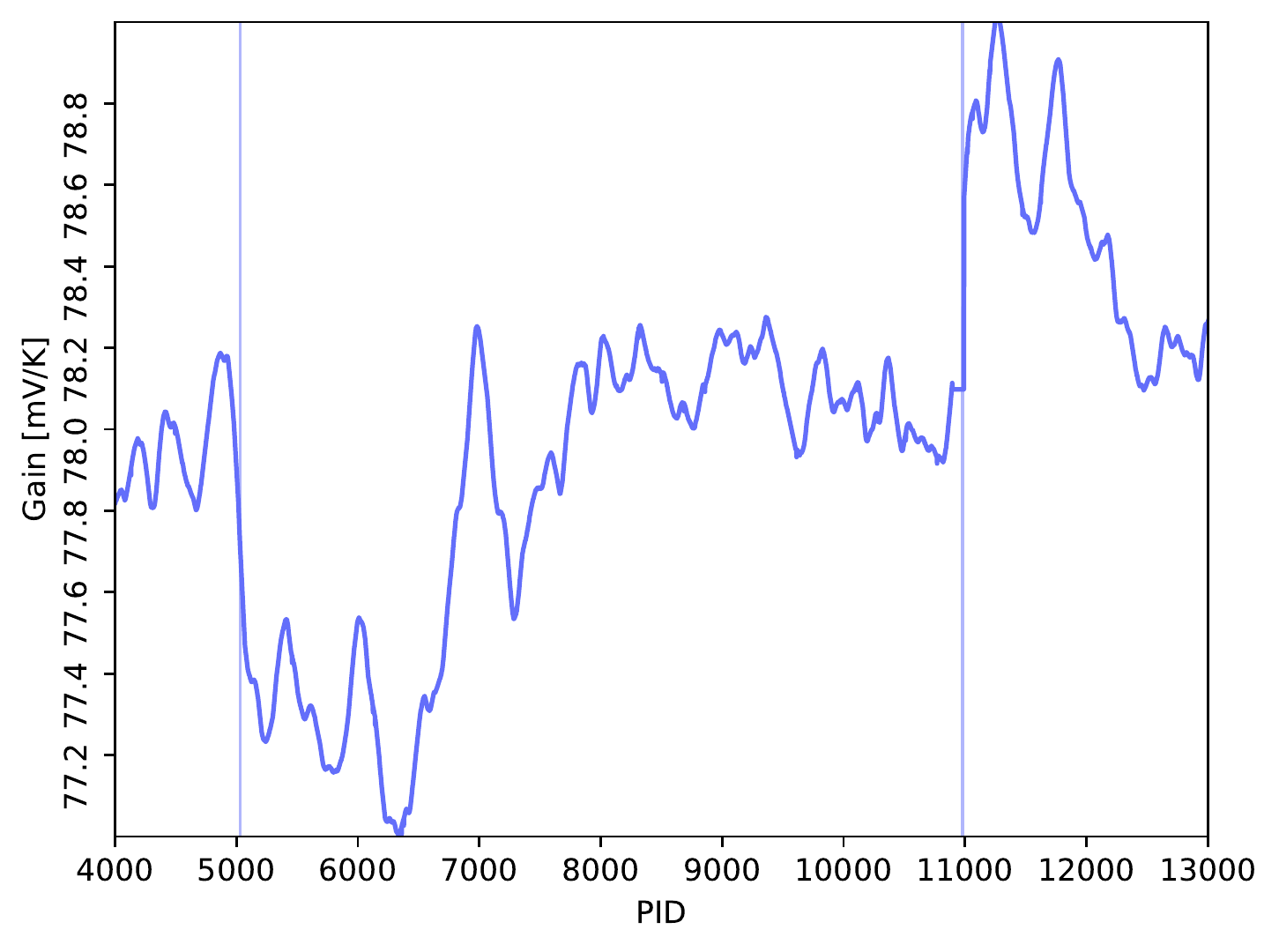}
    \caption{Examples of jumps seen in the gain factors. The line shows the Wiener filter solution for the 27M detector, while the vertical lines show the locations of the jumps.}
  \label{fig:gain_jumps}
\end{figure}

\subsection{Gain posterior distributions}

We start by considering the sampling efficiency of the Monte Carlo
chains produced with the above algorithm in terms of mixing and Markov
chain correlation lengths. Figure~\ref{fig:chains} shows total gain as
a function of Gibbs iteration for some representative radiometers and
PIDs. We see that overall, the chains are stable and mix well.

A more quantitive confirmation of the good mixing of the chains can be seen in
Fig.~\ref{fig:corrlengths}, which shows the correlation lengths across all
PIDs (black line is the estimated mean correlation length, whereas the blue
bands show the estimated standard deviation). All detectors exhibit
correlation coeffiecients less than 10\% after only a few samples.

In Fig.~\ref{fig:allpid_gains}, we show relative differences between
the last sample of the chain and the first drawn sample (red line),
and we compare that to the similar relative difference with the
second-to-last sample in the chain (blue line), as well as an
in-between chain comparison (green line). We see that even the first
sample of the chain is as close to the final solution as the
next-to-last sample, and the conditional burn-in period with respect
to the gain does not significantly affect our results. Long-term
burn-in is caused indirectly through correlations with other
parameters in the system. Because of these external correlations, we
follow \citet{bp01}, and omit the first 200 samples when presenting the
final gain estimates.

In Figs.~\ref{fig:gaincomp_030}--\ref{fig:gaincomp_070}, we compare
the gain factors derived by \BP, \Planck\ DR4 and \Planck\ 2018 for each
detector. For \BP, the widths of each curve represent 1\,$\sigma$
posterior confidence regions as evaluated directly from the Gibbs
chains (after omitting 200 samples for burn-in), while for the other
two solutions we only show final best-fit estimates.

Overall, the largest differences between \BP\ and the other two pipelines are
observed in the 30\,GHz channel. In particular, we find that the \BP\ gain
model is systematically lower than the 2018 model by about 0.84\,\% and than
DR4 by about 0.67\,\% for this channel, which translates into frequency maps
that are about 0.84\,\% (or 0.67\,\%) \emph{brighter}. We also see that the DR4
and 2018 models agree very well for three of the radiometers, while 28S is an
outlier, for which DR4 is close to \BP. 

The 30\,GHz channel is the most difficult to calibrate
among all three LFI channels, because of its brighter foreground
signal, and the different ways in which the three pipelines handle this fact makes the above-mentioned gain solution differences less surprising: The DR4 process treats this channel separately,
in that this channel is analyzed without priors on polarized foregrounds.
The resulting map is then subsequently used as a spatial
polarization prior for the 44 and 70\,GHz channels
\citep{planck2020-LVII}. In comparison, the 2018 approach also assumes
vanishing CMB polarization during calibration, but this approach make
no distinction between the orbital and Solar dipole with respect to
absolute gain calibration (as both DR4 and \BP\ do), but rather
assumes that the fitted foreground model is sufficiently accurate. In
contrast, the \BP\ pipeline does not treat the 30\,GHz channel
differently in any way, and also does not assume that the CMB
polarization signal vanishes (except for the single quadrupole mode,
as discussed in Sect.~\ref{sec:quadrupole}). Instead, it uses
\WMAP\ information to support the foreground modelling, and to
constrain the poorly measured modes in LFI. Overall, these
algorithmic differences lead to the observed deviations between the
various solutions.

The relative differences are smaller for 44 and 70\,GHz: $-0.04\,\%$  (44\,GHz) and $-0.64\,\%$ (70\,GHz) between \BP\ and the \Planck\ 2018 model, and $0.12\,\%$ (44\,GHz) and -$0.03\,\%$ (70\,GHz) between \BP\ and \Planck\ DR4. There is generally good agreement between the three pipelines for these two channels, although \Planck\ 2018 is generally a higher absolute calibration than for the two others. Differences between \BP\ and the two other pipelines are expected due to the joint nature of the \BP\ approach, and the different ways in which the smoothing of the solutions are performed -- \BP\ uses the Wiener filter smoothing method, \Planck\ 2018 uses boxcar averaging, and DR4 does not smooth the solution after estimation at all.

\subsection{Effects of different gain smoothing approaches on correlated noise stripes}

\begin{figure*}[p]
  \center
  \includegraphics[width=0.27\linewidth]{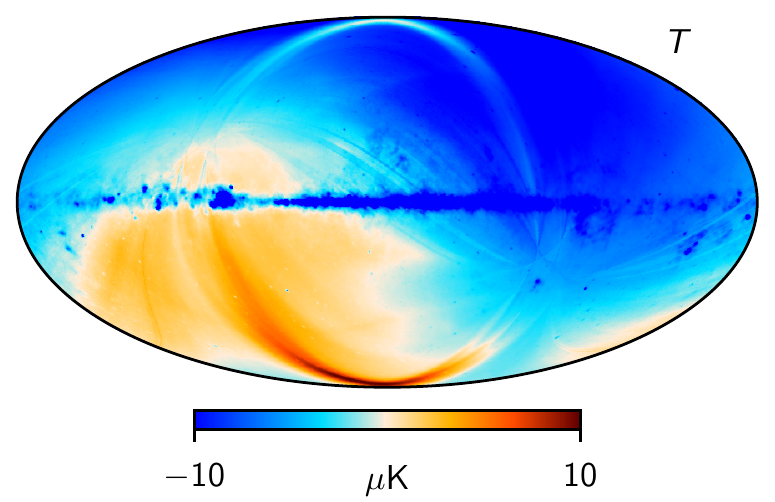}
  \includegraphics[width=0.27\linewidth]{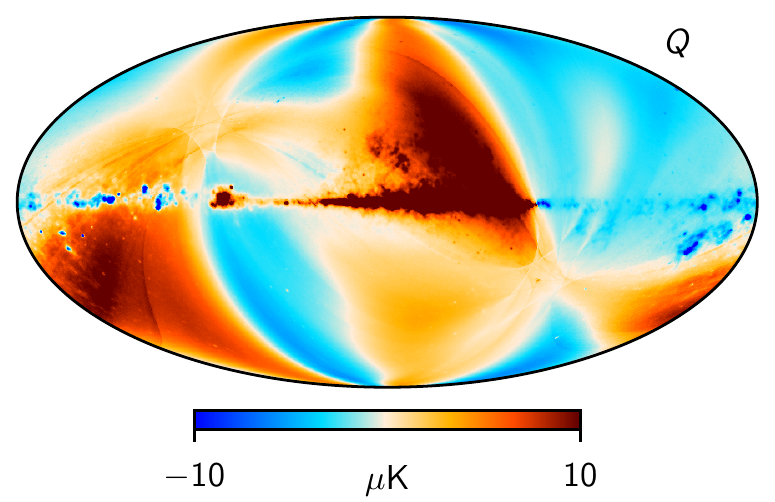}
  \includegraphics[width=0.27\linewidth]{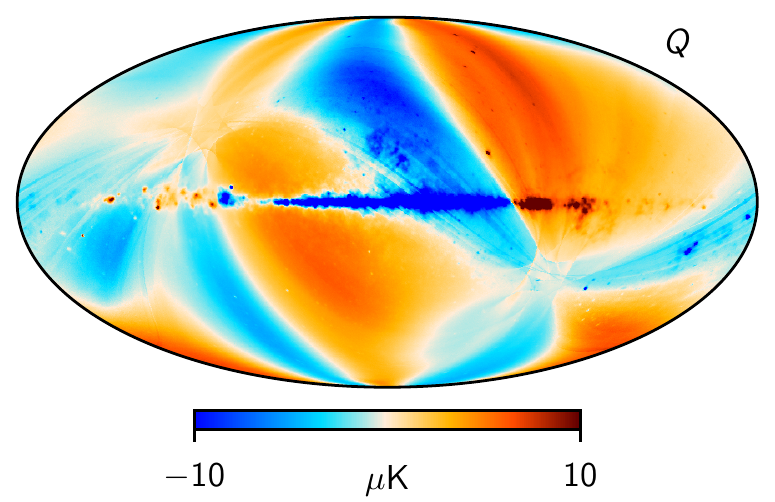}
    \caption{Gain residual template for the LFI 30\,GHz channel, produced in the \Planck\ 2018 analysis through manual iteration between calibration, mapmaking and component separation \citep{planck2016-l02}.
    }
  \label{fig:gain_template}

  \vspace*{1cm}
  
  \center
  \includegraphics[width=0.27\linewidth]{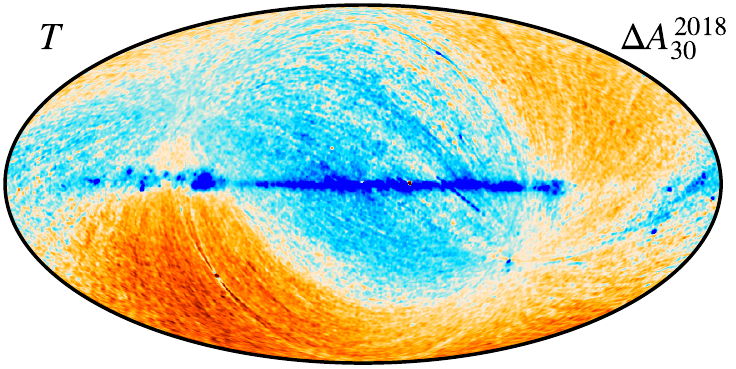}
  \includegraphics[width=0.27\linewidth]{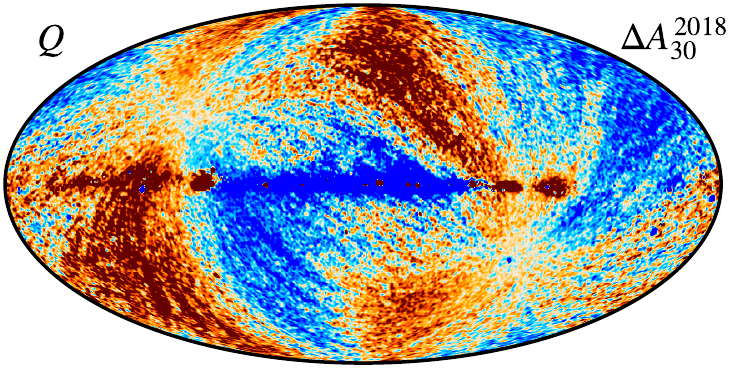}
  \includegraphics[width=0.27\linewidth]{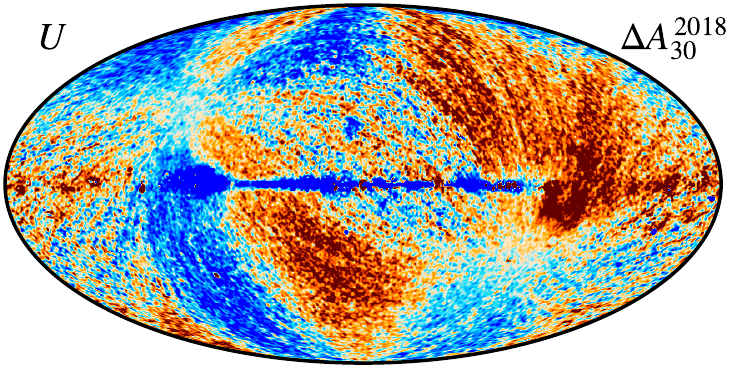}\\
  \includegraphics[width=0.27\linewidth]{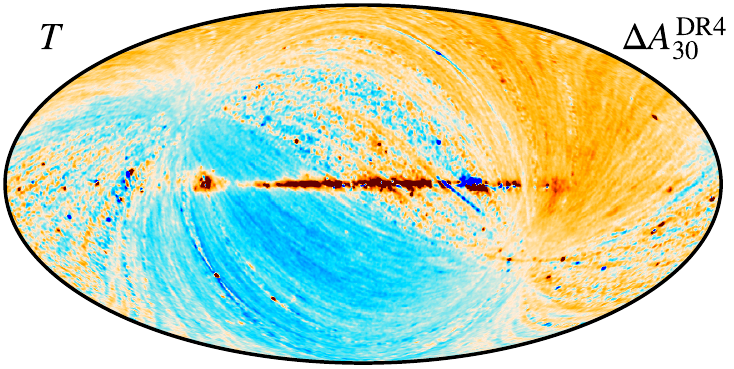}
  \includegraphics[width=0.27\linewidth]{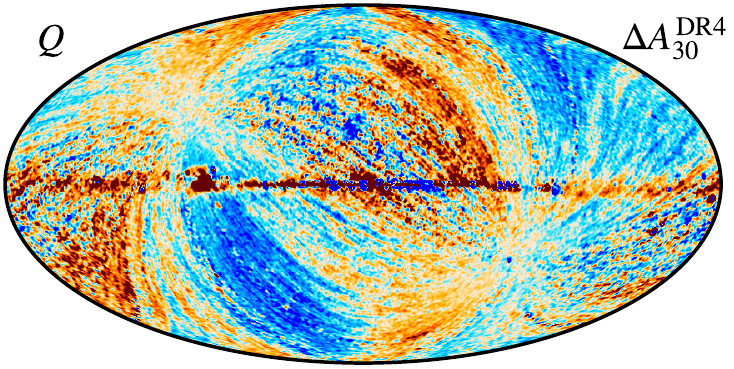}
  \includegraphics[width=0.27\linewidth]{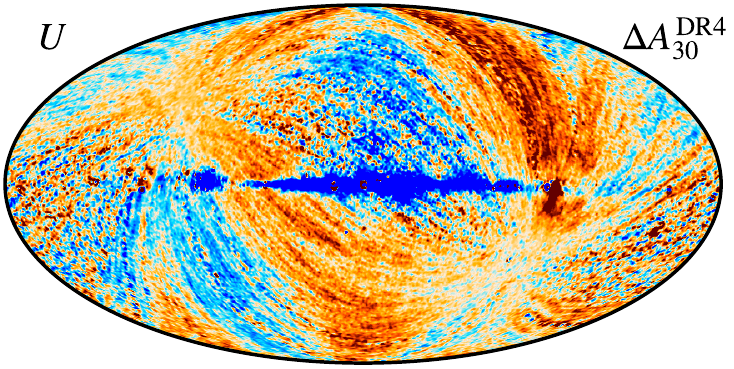}\\
  \vspace*{5mm}
  \includegraphics[width=0.27\linewidth]{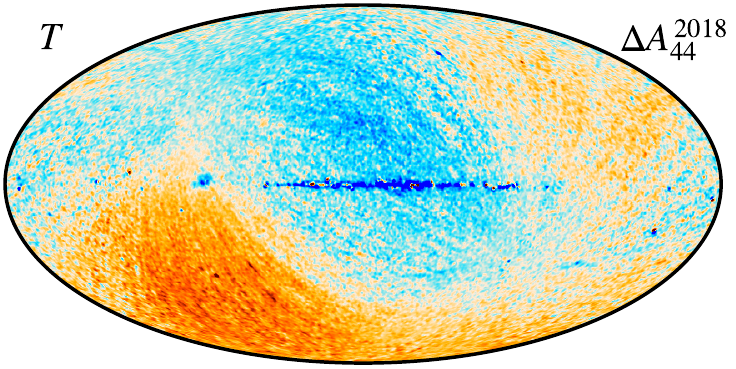}
  \includegraphics[width=0.27\linewidth]{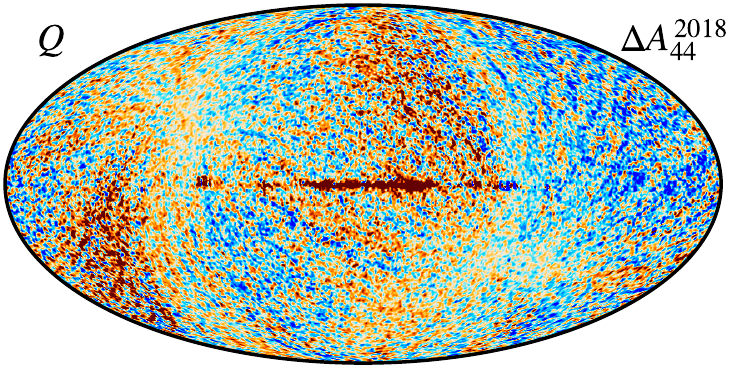}
  \includegraphics[width=0.27\linewidth]{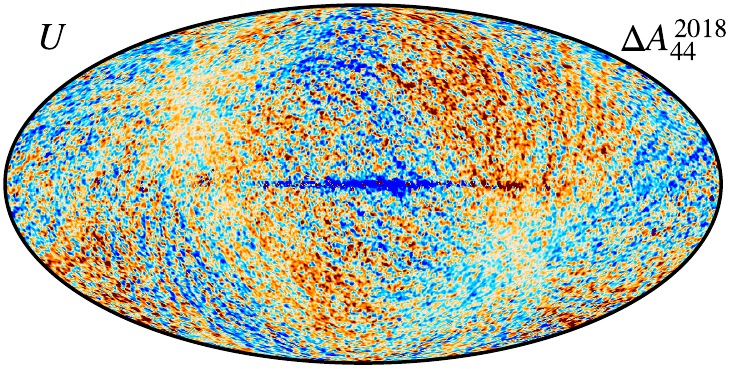}\\
  \includegraphics[width=0.27\linewidth]{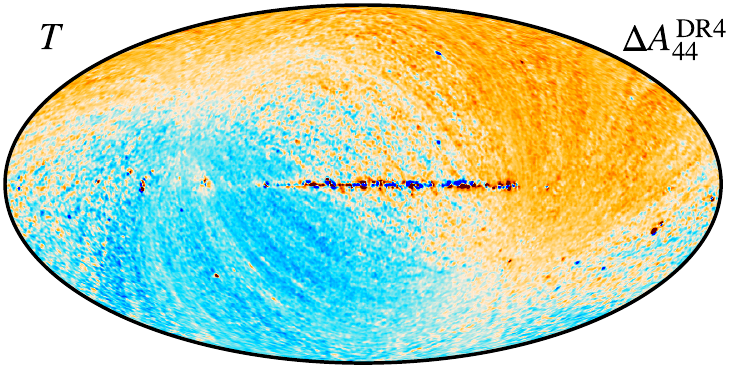}
  \includegraphics[width=0.27\linewidth]{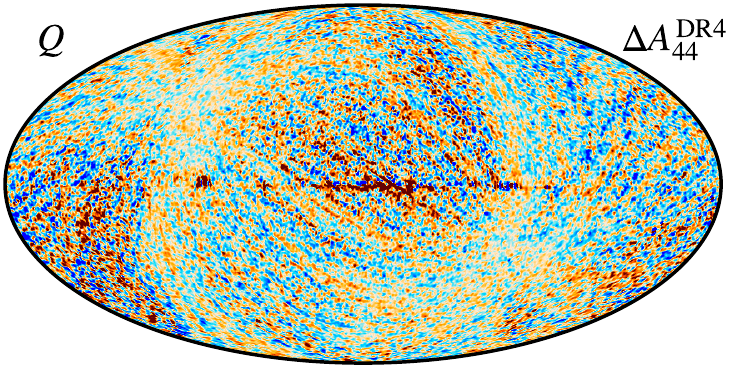}
  \includegraphics[width=0.27\linewidth]{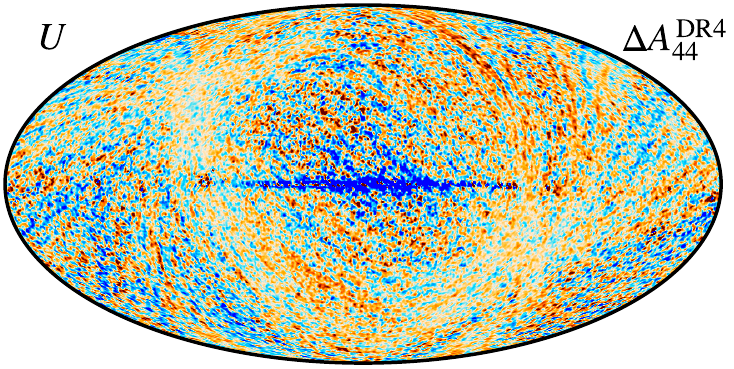}\\
  \vspace*{5mm}
  \includegraphics[width=0.27\linewidth]{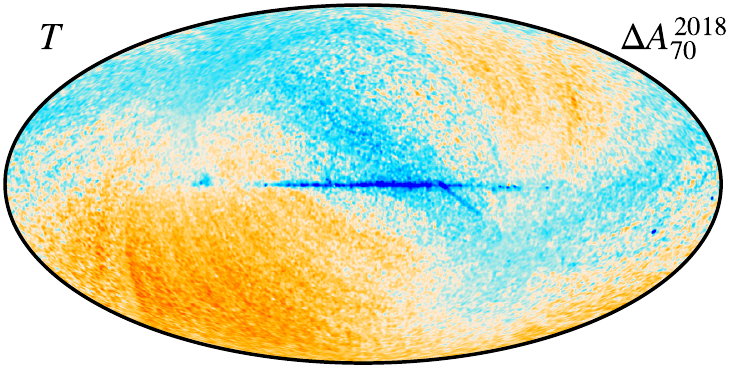}
  \includegraphics[width=0.27\linewidth]{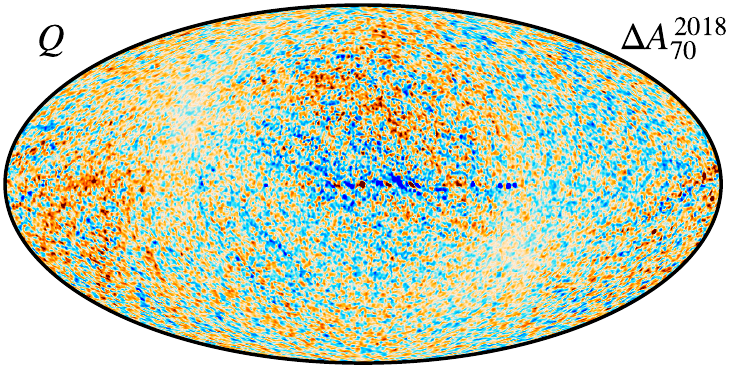}
  \includegraphics[width=0.27\linewidth]{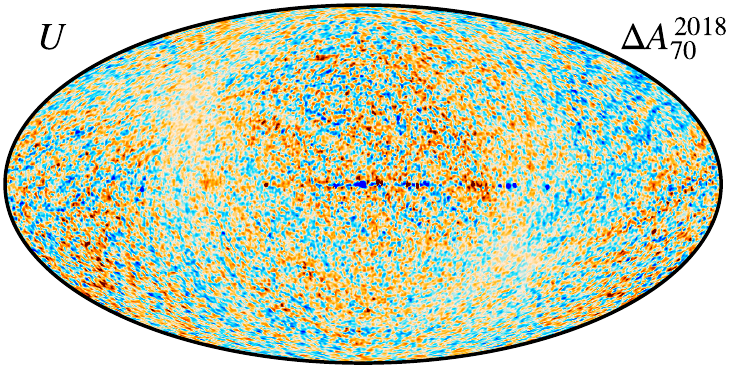}\\
  \includegraphics[width=0.27\linewidth]{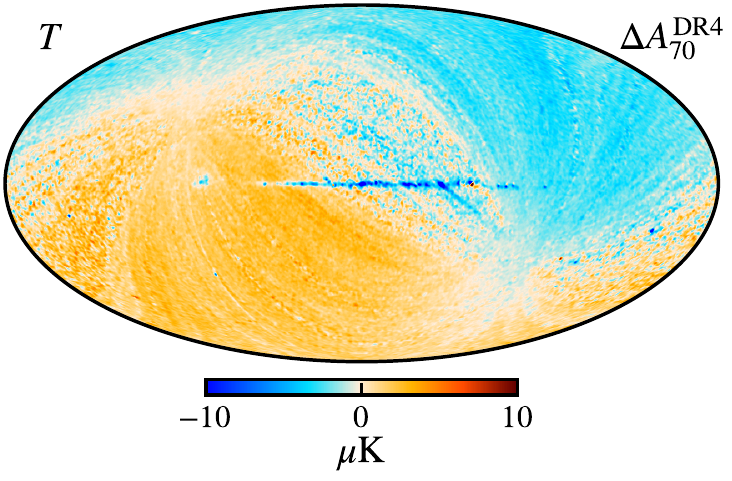}
  \includegraphics[width=0.27\linewidth]{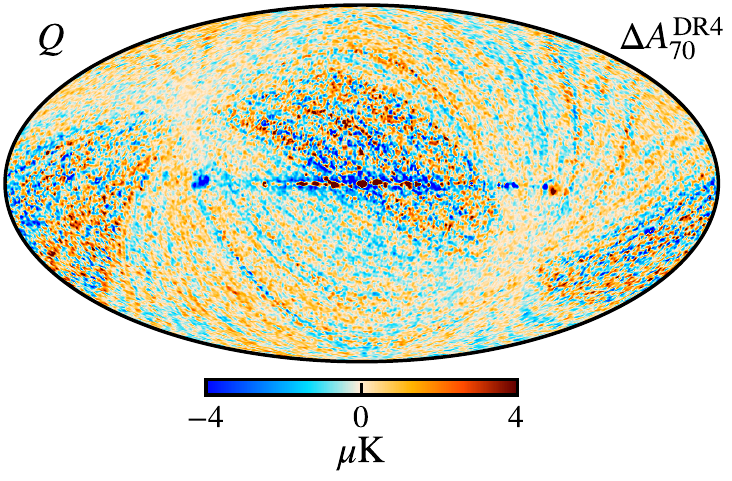}
  \includegraphics[width=0.27\linewidth]{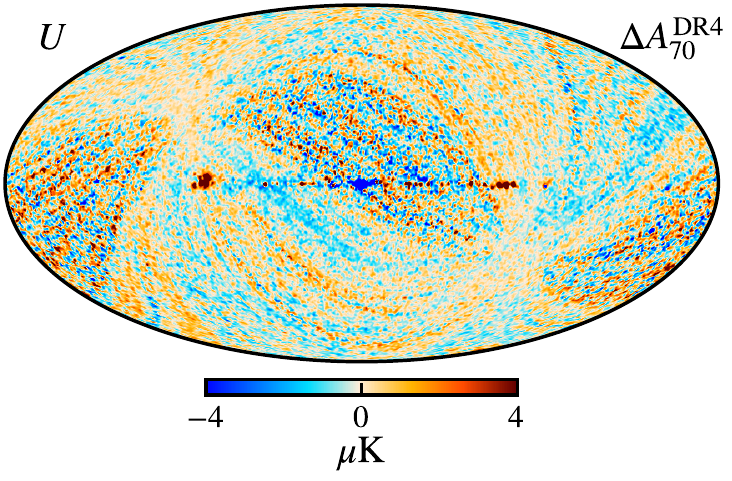}\\
  \caption{Differences between \BP\ and \Planck\ 2018 or \Planck\ DR4 frequency
    maps, smoothed to a common angular resolution of $2^{\circ}$
    FWHM. Columns show Stokes $T$, $Q$, and $U$ parameters,
    respectively, while rows show pairwise differences with respect
    to the pipeline indicated in the panel labels. A constant offset
    has been removed from the temperature maps, while all other modes
    are retained. The 2018 maps have been scaled by their respective
    beam normalization prior to subtraction. Reproduced from \citet{bp10}.
  }\label{fig:freqdiff}

\end{figure*}

Since the Wiener filtering approach used in this paper has not previously been used for smoothing gain estimates, we compared it with a more conventional boxcar smoothing approach with similar smoothing windows as used by \Planck\ DR4 -- that is, smoothing windows that dynamically change with the dipole amplitude so as to make the smoothing length shorter in areas of higher signal-to-noise (i.e. dipole maxima).

When comparing the results of these two approaches, we found that with boxcar averaging, the gain solutions behave more like the original \Planck\ 2018 solutions than is the case with Wiener filtering. However, we also found that boxcar smoothing results in correlated noise solutions with strong ``stripes'' in the binned polarization maps, especially Stokes $Q$. Reducing the window sizes would mitigate this to some degree (although it introduced other issues, like gain spikes in the low signal-to-noise regime), and we therefore find it likely that the correlated noise stripes are related to an ``over-smoothing'' of the gain solution.

With the Wiener filter process, we are, modulo the exact form of the prior covariance matrix, smoothing the data in an optimal way -- in high signal-to-noise areas, the signal is allowed to determine the solution, and in low signal-to-noise areas, the prior ensures that the solution does not become nonphysical. In Figure~\ref{fig:corrstripes}, we show a comparison of the effect of these two approaches on the Stokes $Q$ map. In this figure, the only difference in the underlying sampling algorithm is the choice of boxcar smoothing and Wiener filtering, and it is reassuring to find that Wiener filtering, in line with expectations, is apparently able to find smooth gain solutions that avoid the problem of over-smoothing.

\subsection{Gain jumps}
\label{sec:gainjumps}
As noted from the beginning of the \Planck\ experiment \citep[see, e.g.,][]{planck2013-p05}, the physical gain of the instrument exhibits several sharp jumps. These jumps are related to changes in the thermal environment of the instrument -- an example of such an event is the turning off of the 20\,K sorption cooler. Not all of the events are well understood, and can mainly be traced after an initial gain estimation.

Such jumps must be accounted for in any gain estimation approach that isn't purely data-driven -- i.e. if we are applying priors in any way that set up expected correlated behaviors between the gain factors over the mission. We find that indeed the Wiener filter approach is able to account for such jumps without any extra hard-coding (see Fig.~\ref{fig:gain_jumps}, where a Wiener filter solution applied globally to the whole PID range is shown along with two jumps), another advantage of this smoothing method compared to the boxcar approach, where such jumps must be specified in the code and explicitly excluded from the averaging.

\subsection{Comparison with external data}
\label{sec:external}

To understand the combined impact of the various gain model
differences discussed above, it is useful to compare the final
\BP\ frequency maps with externally processed observations, both from
\WMAP\ and \Planck. In this respect, we note that both the
\Planck\ 2018 and \WMAP\ data sets are associated with sets of
correction templates that track known systematic effects (or poorly
measured modes) in the respective sky maps. For the \Planck\ 2018
30\,GHz channel, this template is shown in
Fig.~\ref{fig:gain_template}. As discussed by \citet{planck2016-l02},
this template was produced by iterating between calibration and
component separation, and therefore traces uncertainties in the gain
model due to foreground uncertainties. Furthermore, due to limited
time, only four full iterations of this type were completed for the
\Planck\ 2018 analysis, and one must therefore expect that there are
still residuals of this type present in the final sky maps at some
level.

With this in mind, we show in Fig.~\ref{fig:freqdiff} \BP--DR4 and
\BP--\Planck\ 2018 difference maps for Stokes $I$, $Q$, and $U$
(columns), for all three \lfi\ frequencies (rows). The first, third
and fifth rows show differences with respect to \Planck\ 2018, while
the second, fourth and sixth rows show differences with respect to
DR4. Several features in these difference maps are interesting from
the calibration perspective. Starting with the DR4 temperature
difference maps, we see that all three channels are dominated by a
clean dipole-like residual aligned with the Solar CMB dipole. This
shows that the \BP\ and DR4 temperature maps are morphologically
very similar, but have different absolute calibration. We also see
that the temperature map difference between \BP\ and DR4 exhibits
a flip in the dipole direction going from 30 and 44 to 70\,GHz. This
sign change is consistent with the differences in the calibration
factors between 44 and 70\,GHz reported in Table~10 in
\citet{planck2020-LVII}, finding a difference of 0.31\,\% between the
absolute calibration of the 44 and 70\,GHz channels. Since the CMB
Solar dipole has an amplitude of about 3360\muK, this relative
difference translates into an absolute temperature difference of
roughly 10~$\muK$ in the observed sky signal, which is fully
consistent with the dipole differences we see in
Fig.~\ref{fig:freqdiff}. In comparison, the \Planck\ 2018 
vs BP difference maps in temperature (line 1, 3, and 5) show a more prominent
quadrupolar structure with a morphology that might resemble the effect of
bandpass mismatch leakage \citep{planck2014-a12,planck2020-LVII}.

\begin{figure*}[p]
    \center
    \includegraphics[width=0.495\linewidth]{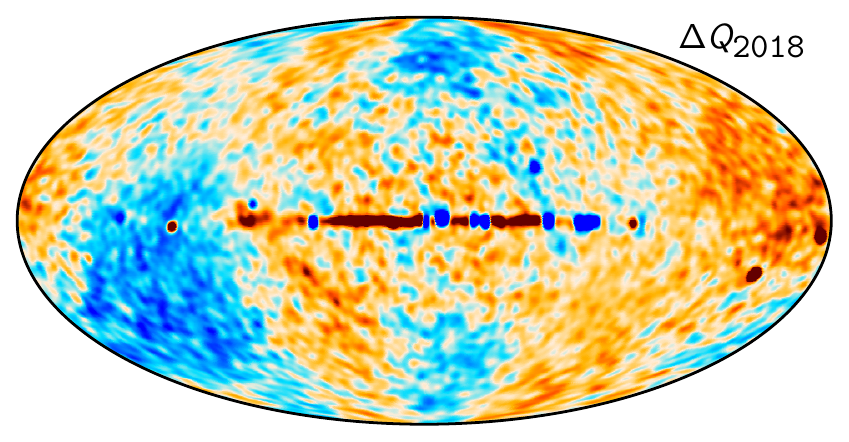}
    \includegraphics[width=0.495\linewidth]{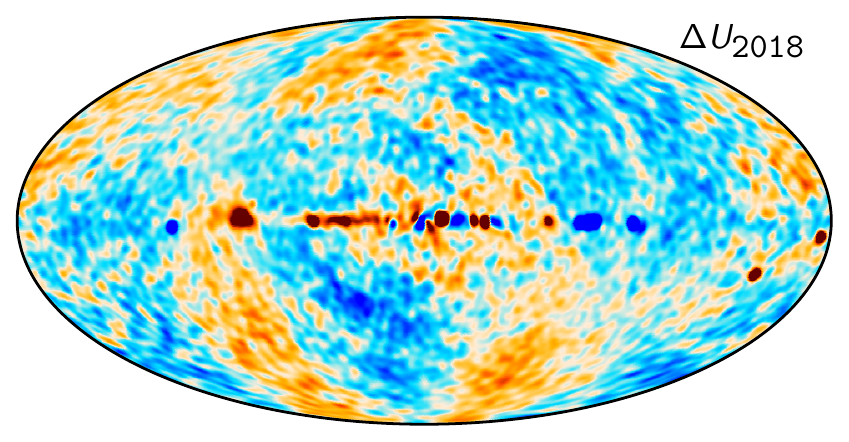}\\
    \includegraphics[width=0.495\linewidth]{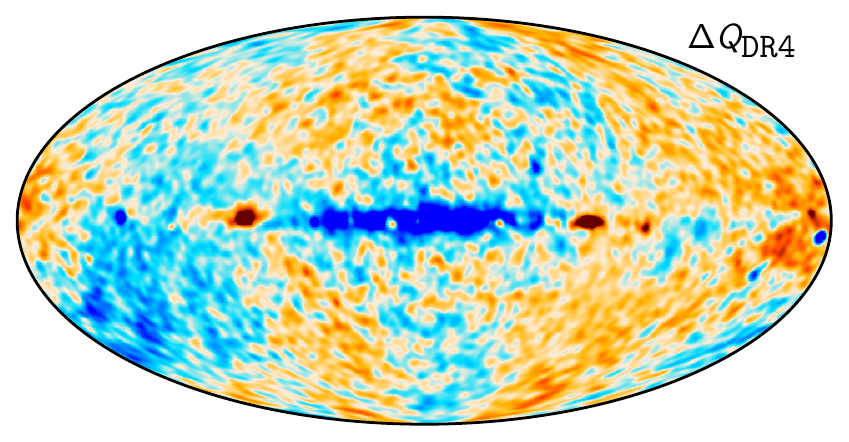}
    \includegraphics[width=0.495\linewidth]{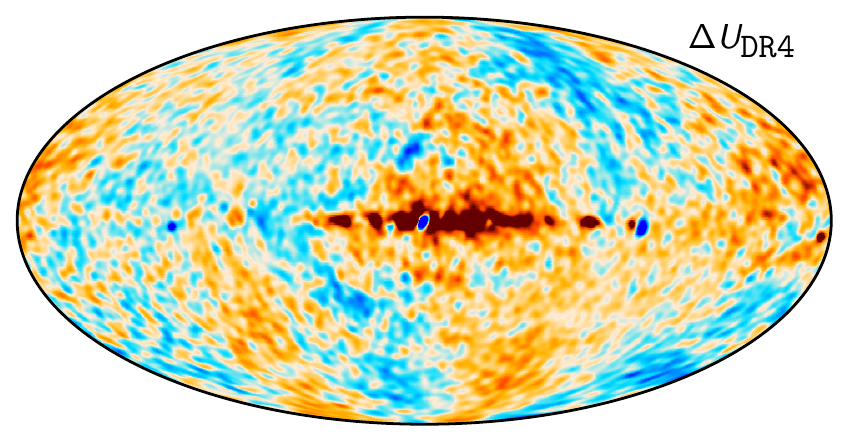}\\
    \includegraphics[width=0.495\linewidth]{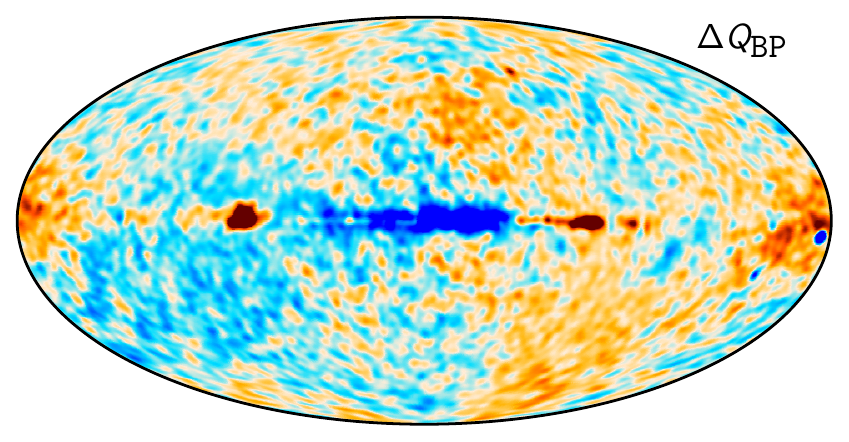}
    \includegraphics[width=0.495\linewidth]{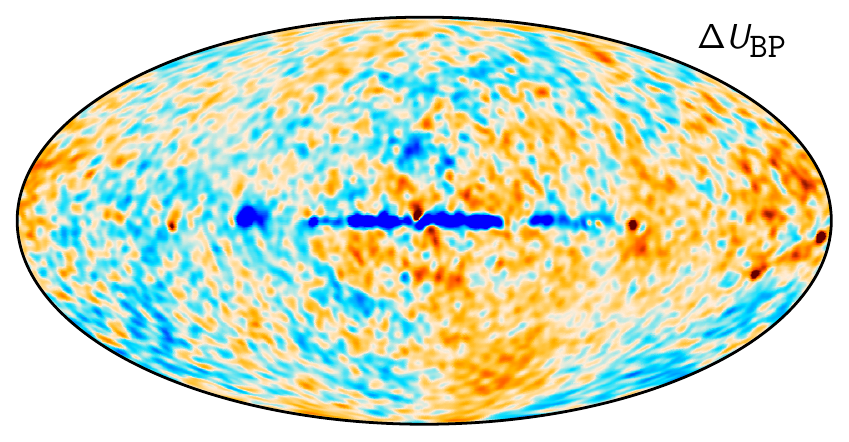}\\
    \includegraphics[width=0.495\linewidth]{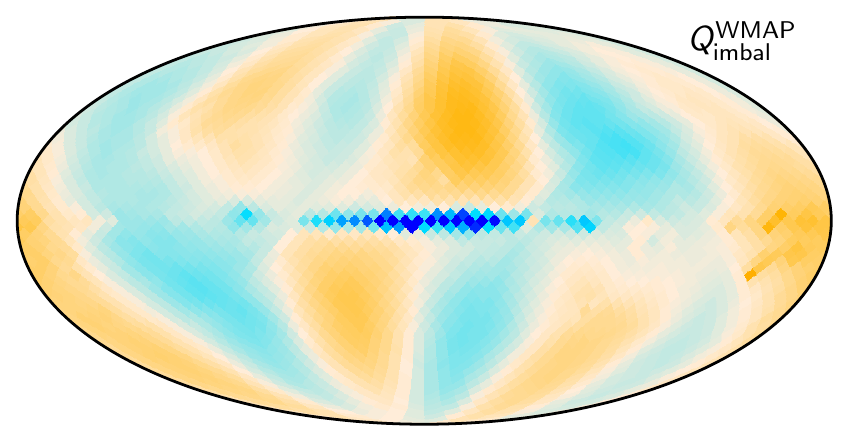}
    \includegraphics[width=0.495\linewidth]{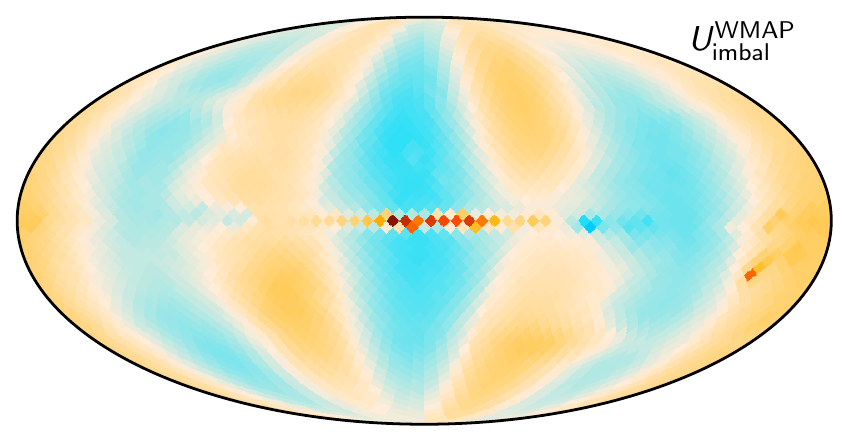}\\    
    \includegraphics[width=0.40\linewidth]{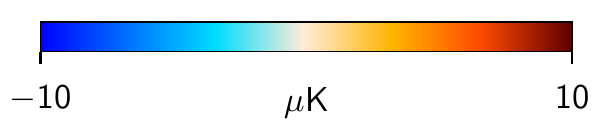}
    \caption{Difference maps between the \Planck\ 30\,GHz and \WMAP\ $K$-band maps for \Planck\ 2018 (\emph{first row}), \Planck\ DR4 (\emph{second row}), and \BP\ (\emph{third row}). All maps have been smoothed to a common angular resolution of $3^{\circ}$ FWHM before evaluating the differences. The \WMAP\ $K$-band map has been scaled by a factor of 0.495 to account for different center frequencies, assuming a synchrotron spectral index of $\beta_{\mathrm{s}}=-3.1$. The bottom row shows one of the \WMAP\ $K$-band transmission imbalance templates discussed by \citet{jarosik2007}, which accounts for known poorly measured modes in the \WMAP\ data. }
    \label{fig:diff_30_k}
\end{figure*}

For polarization, the most striking differences are seen in the
30\,GHz channel, for which variations at the 4\muK\ level are seen
over large fractions of the sky. Furthermore, these residuals
correlated very closely with the \Planck\ 2018 gain template shown in
Fig.~\ref{fig:gain_template}, suggesting that they are indeed caused
by foreground-induced gain residuals. The same patterns are also seen
in the DR4 difference maps, but with a notably lower level.

For the 44\,GHz maps, we note that the Stokes $Q$ difference
maps show correlated noise stripes similar to those highlighted in
the top figure in Fig.~\ref{fig:corrstripes}. However, we also note that these
structures have different amplitudes in \Planck\ DR4 and \Planck\ 2018,
and these stripes are therefore present in at least one of the other
pipelines as well, and possibly both. 

Figure~\ref{fig:diff_30_k} shows a similar comparison between the
various \Planck\ 30\,GHz maps and the \WMAP\ \emph{K}-band channel
\citep{bennett2012}. In this case, all maps have been smoothed to a
common angular resolution of 3$^\circ$ FWHM, and the $K$-band map has
been scaled by a factor of 0.495 to account for the different center
frequencies of the two maps while adopting a synchrotron-like spectral
index of $\beta_{\mathrm{s}}=-3.1$. From top to bottom, the first
three rows show difference maps with respect to \Planck\ 2018, \Planck\ DR4,
and \BP.

Overall, we see a clear progression in agreement with respect to
\WMAP\ $K$-band, in the sense that \BP\ shows smaller residuals than
\Planck\ DR4, which in turn shows smaller residuals than
\Planck\ 2018. Furthermore, we note that the strong residuals traced
by the LFI gain template in Fig.~\ref{fig:gain_template} are most
pronounced in the \Planck\ 2018 map.

At the same time, we also observe significant coherent large-scale
features in the difference map between \BP\ and $K$-band. To at least
partially understand these, we show the \WMAP\ transmission imbalance
templates derived by \citet{jarosik2007} in the bottom row of
Fig.~\ref{fig:freqdiff}. These templates trace poorly measured modes
due to the differential nature of the \WMAP\ instrument. Although
corrections for this effect are applied to the final $K$-band sky
maps, the uncertainty on the template amplitudes is estimated to
20\,\%. Considering the tight correlation between the
\BP--\WMAP\ difference map and the transmission imbalance template, it
seems clear that at least a significant fraction of the remaining
residual may be explained in terms of this effect. Of course, this
also suggests that a future joint analysis between \Planck\ and
\WMAP\ in time-domain will be able to constrain the
\WMAP\ transmission imbalance parameters to much higher precision, and
\Planck\ data can thereby be used to break an important internal
degeneracy in \WMAP. As reported by \citet{bp17}, this work has
already started, but a full exploration of time-ordered \WMAP\ data is
outside the scope of the current analysis. We also emphasize that the
current \BP\ analysis only uses low-resolution \WMAP\ polarization data
for which a full covariance matrix is available, and these modes are
appropriately down-weighted in those matrices.

\section{Conclusions}
\label{sec:conclusions}

We have presented the \BP\ approach to gain calibration within the
larger \commander\ Gibbs sampling framework. This framework relies
directly on the Solar and orbital dipoles for relative and absolute
calibration, respectively, and accounts for astrophysical foreground
and instrumental systematics through global modelling.

One critically important difference with respect to previous
\Planck\ LFI analysis efforts is the fact that we actively use
external data to break internal \Planck\ degeneracies, and in
particular \WMAP\ observations. This significantly alleviates the need
for imposing strong algorithmic priors during the calibration
process. Most notably, while both the \Planck\ 2018 and
\Planck\ DR4 pipelines assumed CMB polarization to be negligible on all
angular scales during the calibration phase, we only assume that the
CMB quadrupole is negligible. The reason we still make this assumption
is that the \Planck\ scanning strategy renders the CMB quadrupole very
nearly perfectly degenerate with the CMB Solar dipole coupled to
subtle gain fluctuations; a hypothetical future and well-designed
satellite mission should not require this prior, as long as its
scanning strategy modulates the CMB dipole on sufficiently short
time-scales and with good cross-linking.

Overall, we find good agreement between the \BP\ and previous gain
models. The biggest differences are observed in the LFI 30\,GHz
channel, with gain variations of 0.84\,\%  between \Planck\ 2018 and \BP. These differences
result in subtle but significant temperature and polarization
residuals. When comparing these with external \WMAP\ $K$-band
observations, it seems clear that the \BP\ LFI maps are generally
cleaner than previous renditions with respect to gain residuals. At
the same time, we emphasize that these differences are also consistent
with previously published error estimates, as presented by the
\Planck\ 2018 and \Planck\ DR4 teams themselves. For instance, the
morphology of the \Planck\ 2018 polarization residuals matches
previously published LFI DPC gain residual templates
\citep{planck2016-l02}, and the DR4 absolute calibration
differences are fully consistent with internal DR4 estimates
\citep{planck2020-LVII}. These results are thus neither novel nor
surprising, but they simply highlight the inherent advantages of
global analysis, using complementary data sets to break internal
degeneracies.

Finally, we note that even though the procedures outlined in this
paper have been aimed at modelling the \lfi\ detectors, there is
nothing about the data model or methodology that is unique for
\lfi. The method should be directly applicable for other data sets and
experiments as well, and, indeed, a preliminary \WMAP\ analysis is
already underway \citep{bp17}.

\input{BP_acknowledgments.tex}

\bibliographystyle{aa}

\bibliography{Planck_bib,BP_bibliography}

\end{document}

%% file: Planck.tex
\def\setsymbol#1#2{\expandafter\def\csname #1\endcsname{#2}}
\def\getsymbol#1{\csname #1\endcsname}

\def\Planck{\textit{Planck}}

\newbox\tablebox    \newdimen\tablewidth
\def\leaderfil{\leaders\hbox to 5pt{\hss.\hss}\hfil}
\def\tablenote#1 #2\par{\begingroup \parindent=0.8em
    \abovedisplayshortskip=0pt\belowdisplayshortskip=0pt
    \noindent
    $$\hss\vbox{\hsize\tablewidth \hangindent=\parindent \hangafter=1 \noindent
    \hbox to \parindent{$^#1$\hss}\strut#2\strut\par}\hss$$
    \endgroup}

\def\L2{\ifmmode L_2\else $L_2$\fi}

\def\DeltaT{\ifmmode \Delta T\else $\Delta T$\fi}
\def\deltat{\ifmmode \Delta t\else $\Delta t$\fi}
\def\fknee{\ifmmode f_{\rm knee}\else $f_{\rm knee}$\fi}
\def\Fmax{\ifmmode F_{\rm max}\else $F_{\rm max}$\fi}
\def\solar{\ifmmode{\rm M}_{\mathord\odot}\else${\rm M}_{\mathord\odot}$\fi}
\def\Msolar{\ifmmode{\rm M}_{\mathord\odot}\else${\rm M}_{\mathord\odot}$\fi}
\def\Lsolar{\ifmmode{\rm L}_{\mathord\odot}\else${\rm L}_{\mathord\odot}$\fi}
\def\inv{\ifmmode^{-1}\else$^{-1}$\fi}
\def\mo{\ifmmode^{-1}\else$^{-1}$\fi}
\def\sup#1{\ifmmode ^{\rm #1}\else $^{\rm #1}$\fi}
\def\expo#1{\ifmmode \times 10^{#1}\else $\times 10^{#1}$\fi}
\def\,{\thinspace}
\def\lsim{\mathrel{\raise .4ex\hbox{\rlap{$<$}\lower 1.2ex\hbox{$\sim$}}}}
\def\gsim{\mathrel{\raise .4ex\hbox{\rlap{$>$}\lower 1.2ex\hbox{$\sim$}}}}

\def\simprop{\mathrel{\raise .4ex\hbox{\rlap{$\propto$}\lower 1.2ex\hbox{$\sim$}}}}
\def\deg{\ifmmode^\circ\else$^\circ$\fi}
\def\pdeg{\ifmmode $\setbox0=\hbox{$^{\circ}$}\rlap{\hskip.11\wd0 .}$^{\circ}
          \else \setbox0=\hbox{$^{\circ}$}\rlap{\hskip.11\wd0 .}$^{\circ}$\fi}
\def\arcs{\ifmmode {^{\scriptstyle\prime\prime}}
          \else $^{\scriptstyle\prime\prime}$\fi}
\def\arcm{\ifmmode {^{\scriptstyle\prime}}
          \else $^{\scriptstyle\prime}$\fi}
\newdimen\sa  \newdimen\sb
\def\parcs{\sa=.07em \sb=.03em
     \ifmmode \hbox{\rlap{.}}^{\scriptstyle\prime\kern -\sb\prime}\hbox{\kern -\sa}
     \else \rlap{.}$^{\scriptstyle\prime\kern -\sb\prime}$\kern -\sa\fi}
\def\parcm{\sa=.08em \sb=.03em
     \ifmmode \hbox{\rlap{.}\kern\sa}^{\scriptstyle\prime}\hbox{\kern-\sb}
     \else \rlap{.}\kern\sa$^{\scriptstyle\prime}$\kern-\sb\fi}
\def\ra[#1 #2 #3.#4]{#1\sup{h}#2\sup{m}#3\sup{s}\llap.#4}
\def\dec[#1 #2 #3.#4]{#1\deg#2\arcm#3\arcs\llap.#4}
\def\deco[#1 #2 #3]{#1\deg#2\arcm#3\arcs}
\def\rra[#1 #2]{#1\sup{h}#2\sup{m}}

\def\dots{\relax\ifmmode \ldots\else $\ldots$\fi}
\def\WHzsr{\ifmmode $W\,Hz\mo\,sr\mo$\else W\,Hz\mo\,sr\mo\fi}
\def\mHz{\ifmmode $\,mHz$\else \,mHz\fi}
\def\GHz{\ifmmode $\,GHz$\else \,GHz\fi}
\def\mKs{\ifmmode $\,mK\,s$^{1/2}\else \,mK\,s$^{1/2}$\fi}
\def\muKs{\ifmmode \,\mu$K\,s$^{1/2}\else \,$\mu$K\,s$^{1/2}$\fi}
\def\muKRJs{\ifmmode \,\mu$K$_{\rm RJ}$\,s$^{1/2}\else \,$\mu$K$_{\rm RJ}$\,s$^{1/2}$\fi}
\def\muKHz{\ifmmode \,\mu$K\,Hz$^{-1/2}\else \,$\mu$K\,Hz$^{-1/2}$\fi}
\def\MJysr{\ifmmode \,$MJy\,sr\mo$\else \,MJy\,sr\mo\fi}
\def\MJysrmK{\ifmmode \,$MJy\,sr\mo$\,mK$_{\rm CMB}\mo\else \,MJy\,sr\mo\,mK$_{\rm CMB}\mo$\fi}
\def\microns{\ifmmode \,\mu$m$\else \,$\mu$m\fi}

\def\muK{\ifmmode \,\mu$K$\else \,$\mu$\hbox{K}\fi}
\def\microK{\ifmmode \,\mu$K$\else \,$\mu$\hbox{K}\fi}
\def\muW{\ifmmode \,\mu$W$\else \,$\mu$\hbox{W}\fi}
\def\kms{\ifmmode $\,km\,s$^{-1}\else \,km\,s$^{-1}$\fi}
\def\kmsMpc{\ifmmode $\,\kms\,Mpc\mo$\else \,\kms\,Mpc\mo\fi}
\providecommand{\sorthelp}[1]{}

%% file: authors_07.tex
%This author list corresponds to \title{Author list for L04\_CMB\_Foregrounds\_Extraction}
%Prepared by M. Lopez-Caniego (Marcos.Lopez.Caniego@sciops.esa.int), ESAC/ESA
%This version is from Thu Jul 12 18:11:48 2018 CET
%\subtitle{There are 152 co-authors in this list}
\newcommand{\oslo}[0]{1}
\newcommand{\triesteB}[0]{2}
\newcommand{\milanoA}[0]{3}
\newcommand{\milanoB}[0]{4}
\newcommand{\milanoC}[0]{5}
\newcommand{\planetek}[0]{6}
\newcommand{\princeton}[0]{7}
\newcommand{\jpl}[0]{8}
\newcommand{\helsinkiA}[0]{9}
\newcommand{\helsinkiB}[0]{10}
\newcommand{\nersc}[0]{11}
\newcommand{\haverford}[0]{12}
\newcommand{\mpa}[0]{13}
\newcommand{\triesteA}[0]{14}
\author{\small
E.~Gjerl{\o}w\inst{\oslo}\thanks{Corresponding author: E.~Gjerl{\o}w; \url{eirik.gjerlow@astro.uio.no}}
\and
\textcolor{black}{H.~T.~Ihle}\inst{\oslo}
\and
S.~Galeotta\inst{\triesteB}
\and
K.~J.~Andersen\inst{\oslo}
\and
\textcolor{black}{R.~Aurlien}\inst{\oslo}
\and
\textcolor{black}{R.~Banerji}\inst{\oslo}
\and
M.~Bersanelli\inst{\milanoA, \milanoB, \milanoC}
\and
S.~Bertocco\inst{\triesteB}
\and
M.~Brilenkov\inst{\oslo}
\and
M.~Carbone\inst{\planetek}
\and
L.~P.~L.~Colombo\inst{\milanoA}
\and
H.~K.~Eriksen\inst{\oslo}
\and
\textcolor{black}{M.~K.~Foss}\inst{\oslo}
\and
C.~Franceschet\inst{\milanoA, \milanoC}
\and
\textcolor{black}{U.~Fuskeland}\inst{\oslo}
\and
M.~Galloway\inst{\oslo}
\and
S.~Gerakakis\inst{\planetek}
\and
\textcolor{black}{B.~Hensley}\inst{\princeton}
\and
\textcolor{black}{D.~Herman}\inst{\oslo}
\and
M.~Iacobellis\inst{\planetek}
\and
M.~Ieronymaki\inst{\planetek}
\and
J.~B.~Jewell\inst{\jpl}
\and
\textcolor{black}{A.~Karakci}\inst{\oslo}
\and
E.~Keih\"{a}nen\inst{\helsinkiA, \helsinkiB}
\and
R.~Keskitalo\inst{\nersc}
\and
G.~Maggio\inst{\triesteB}
\and
D.~Maino\inst{\milanoA, \milanoB, \milanoC}
\and
M.~Maris\inst{\triesteB}
\and
S.~Paradiso\inst{\milanoA, \milanoC}
\and
B.~Partridge\inst{\haverford}
\and
M.~Reinecke\inst{\mpa}
\and
A.-S.~Suur-Uski\inst{\helsinkiA, \helsinkiB}
\and
T.~L.~Svalheim\inst{\oslo}
\and
D.~Tavagnacco\inst{\triesteB, \triesteA}
\and
H.~Thommesen\inst{\oslo}
\and
D.~J.~Watts\inst{\oslo}
\and
I.~K.~Wehus\inst{\oslo}
\and
A.~Zacchei\inst{\triesteB}
}
\institute{\small
Institute of Theoretical Astrophysics, University of Oslo, Blindern, Oslo, Norway\goodbreak
\and
INAF - Osservatorio Astronomico di Trieste, Via G.B. Tiepolo 11, Trieste, Italy\goodbreak
\and
Dipartimento di Fisica, Universit\`{a} degli Studi di Milano, Via Celoria, 16, Milano, Italy\goodbreak
\and
INAF-IASF Milano, Via E. Bassini 15, Milano, Italy\goodbreak
\and
INFN, Sezione di Milano, Via Celoria 16, Milano, Italy\goodbreak
\and
Planetek Hellas, Leoforos Kifisias 44, Marousi 151 25, Greece\goodbreak
\and
Department of Astrophysical Sciences, Princeton University, Princeton, NJ 08544,
U.S.A.\goodbreak
\and
Jet Propulsion Laboratory, California Institute of Technology, 4800 Oak Grove Drive, Pasadena, California, U.S.A.\goodbreak
\and
Department of Physics, Gustaf H\"{a}llstr\"{o}min katu 2, University of Helsinki, Helsinki, Finland\goodbreak
\and
Helsinki Institute of Physics, Gustaf H\"{a}llstr\"{o}min katu 2, University of Helsinki, Helsinki, Finland\goodbreak
\and
Computational Cosmology Center, Lawrence Berkeley National Laboratory, Berkeley, California, U.S.A.\goodbreak
\and
Haverford College Astronomy Department, 370 Lancaster Avenue,
Haverford, Pennsylvania, U.S.A.\goodbreak
\and
Max-Planck-Institut f\"{u}r Astrophysik, Karl-Schwarzschild-Str. 1, 85741 Garching, Germany\goodbreak
\and
Dipartimento di Fisica, Universit\`{a} degli Studi di Trieste, via A. Valerio 2, Trieste, Italy\goodbreak
}

%% file: BP_acknowledgments.tex
\begin{acknowledgements}
  We thank Prof.\ Pedro Ferreira for useful suggestions, comments and
  discussions, and Dr.\ Diana Mjaschkova-Pascual for administrative
  support. We also thank the entire \Planck\ and \WMAP\ teams for
  invaluable support and discussions, and for their dedicated efforts
  through several decades without which this work would not be
  possible. The current work has received funding from the European
  Union’s Horizon 2020 research and innovation programme under grant
  agreement numbers 776282 (COMPET-4; \BP), 772253 (ERC;
  \textsc{bits2cosmology}), and 819478 (ERC; \textsc{Cosmoglobe}). In
  addition, the collaboration acknowledges support from ESA; ASI and
  INAF (Italy); NASA and DoE (USA); Tekes, Academy of Finland (grant
   no.\ 295113), CSC, and Magnus Ehrnrooth foundation (Finland); RCN
  (Norway; grant nos.\ 263011, 274990); and PRACE (EU).
\end{acknowledgements}